\documentclass[aps,showpacs]{revtex4}
\usepackage{graphicx,color}
\newcommand{\ds}{ _{\downarrow}}
\newcommand{\us}{ _{\uparrow}}
\newcommand{\up}{\uparrow}
\newcommand{\down}{\downarrow}

\begin{document}
\draft \title{Interaction beween polarons and analogous effects in polarized Fermi gases}
\author{S. Giraud$^{(a)}$ and R. Combescot$^{(b),(c)}$} 
\address{(a) Institut f\"ur Theoretische Physik, Heinrich-Heine-Universit\"at, D-40225  D\"usseldorf, Germany}
\address{(b) Laboratoire de Physique Statistique, Ecole Normale Sup\'erieure, UPMC  
Paris 06, Universit\'e Paris Diderot, CNRS, 24 rue Lhomond, 75005 Paris, France.}
\address{(c) Institut Universitaire de France}
\date{Received \today}

\begin{abstract}
We consider an imbalanced mixture of two different ultracold Fermi gases, which are strongly interacting.
Calling spin-down the minority component and spin-up the majority component, the limit of small relative
density $x=n\ds /n\us$
is usually considered as a gas of non interacting polarons. This allows to calculate, in the expansion of the total
energy of the system in powers of $x$, the terms proportional to $x$ (corresponding to the binding energy of
the polaron) and to $x^{5/3}$ (corresponding to the kinetic energy of the polaron Fermi sea).
We investigate in this paper terms physically due to an interaction between polarons and which are proportional
to $x^2$ and $x^{7/3}$. We find three such terms. A first one corresponds to the overlap between
the clouds dressing two polarons. The two other ones are due to the modification of the single polaron binding energy
caused by the non-zero density of polarons. The second term is due to the restriction of the polaron momentum by
the Fermi sea formed by the other polarons. The last one results from the modification of 
the spin-up Fermi sea brought by the other polarons. The calculation of all these terms is made at the simplest level
of a single particle-hole excitation. It is performed for all the possible interaction strengths within the stability range
of the polaron. At unitarity the last two terms give a fairly weak contribution while the first one is strong and leads
to a marked disagreement with Monte-Carlo results. The possible origins of this discrepancy are discussed.
\end{abstract}
\maketitle

\section{introduction}
Ultracold atomic gases have proved to be remarkably interesting systems due to the extreme simplicity of their effective interactions, which can indeed be
described in many cases by the mere knowledge of the scattering lengths between the involved atoms. In particular fermionic gases made of two kinds of atoms,
usually two hyperfine states of a same atomic element, have their interactions fully described by the scattering length $a$ between different atoms.
A large number of studies have been devoted to the case of balanced systems where the numbers of the two different atoms are the same \cite{gps}.
More recently attention has also turned toward the case of imbalanced mixtures, where these atom numbers are unequal and which display
very rich physics. They are analogous to systems found in other fields of physics, in particular superconductors in very high magnetic fields, but 
the ability to vary at will the ratio between the two atomic populations makes them especially convenient to explore experimentally.

A particularly interesting situation is the limiting case where one atomic species is very dilute compared to the other one, so that its understanding
reduces to the study of a single fermion in the Fermi sea of the other population. We follow the fairly standard convention of calling this isolated atom
a "spin-down" atom while the Fermi sea is made of "spin-up" atoms and consider only the case where this interaction is attractive. This interaction between the spin-down and spin-up atoms "polarizes" the Fermi sea
and the resulting complex object, a quasiparticle, is quite often called a "polaron". It is characterized by its binding energy $E_b=-\mu \ds$, where $\mu \ds$ is the
chemical potential of the spin-down atom, and also by its effective mass $m^*$ which gives the kinetic energy $p^2/2m^*$ when this quasiparticle
has momentum ${\bf p}$. This polaron has been studied by various theoretical approaches \cite{fred, lrgs,crlc,ps1,ps2,cg}, variational, diagrammatic
and Monte-Carlo, which have been found in very good agreement with experiments \cite{nnj,nnj1,zwi}.

The study of the polaron has been the basic step in the theoretical investigation by Lobo {\it et al} \cite{lrgs} of the first order transition observed for
trapped imbalanced Fermi gases by the Rice and the MIT groups \cite{rimit}. Their study has been performed at unitarity, but it has been extended
later on to the whole BEC-BCS crossover \cite{pg}. They have evaluated the total energy of a normal gas of $n\ds$ polarons
in a Fermi sea of $n\us$ atoms by taking into account not only the binding energy of these polarons, but also the kinetic energy of the polaron Fermi sea, proportional to $x^{5/3}$, where $x=n\ds /n\us$ is the spin-down concentration compared to the spin-up one.
These two ingredients have proved essential to the analysis of experiments
and the quite good agreement between experiments and theory. More generally the approach starting from the dilute limit for the spin-down
population and leading for the total energy of the system to an expansion in powers for the spin-down density $x$
has proved to be extremely fruitful. Our paper is in this line of reasoning.

However Monte-Carlo calculations also performed by Lobo {\it et al} \cite{lrgs} showed a small and progressive departure
for larger concentration $x$ from the energy just given by the sum of the polaron chemical potential and the kinetic energy
of the polaron Fermi sea. Such a departure is naturally expected since, when the polaron concentration is
increased, the dilute regime, where each polaron can be considered as isolated, breaks down and one has
to consider the contribution coming from the interaction between polarons. Once the properties of the isolated polaron is under control, this interaction between polarons is naturally the next problem of
theoretical interest. But it is also of
practical interest in order to have a better understanding of the agreement between theory and
experiment. This is the basic problem we will consider in the present paper. We note that Nishida \cite{nis} as well as Patton and Sheehy \cite{pash}
have investigated the quite interesting question of a possible instability coming from polaron interaction
and leading to the p-wave superfluidity of the polaron gas. However we will restrict ourselves in this paper to the calculation
of the total interaction and we will not proceed to an analysis of the partial wave components of the effective interaction.

More precisely this analysis in terms of polarons is valid only in the dilute regime of small $x$, strictly speaking in the
limit $x \to 0$. In this regime we can meaningfully expand the total energy of the system in powers of $x$. The validity of this power
expansion is only guaranteed for small $x$, although one may hope, as it happens quite often, that it extends actually to fairly larger values of $x$
including for example the value $x_c=0.44$ where Lobo {\it et al} have found a first order transition at unitarity.
The binding energy of the polaron gives in the total energy a term proportional to $x$ and the kinetic energy of the polaron Fermi sea is
proportional to $x^{5/3}$. Hence we are interested in terms with next higher power. These will turn out to be terms
proportional to $x^2$ and $x^{7/3}$. It is worth noting that, although these exponents are indeed higher than the $5/3$
arising for the polaron Fermi sea kinetic energy, the corresponding terms are in practice quite relevant for quantitative
evaluations since, in order to neglect them, one would need to have typically $x^{1/3} \sim 0.1$ which implies extremely
small values of $x \sim 10^{-3}$, that is much smaller than the typical values $x_c$ for which the first order transition has been found
at unitarity.

The most obvious physical origin of the interaction between two polarons is the overlap of the spin-up clouds surrounding the 
spin-down fermions, that is the overlap between the polarized regions forming the polarons. This makes each of these
clouds less "perfect" than for an isolated polaron. This reduces the binding energy of each polaron, which implies
a repulsive interaction between these polarons. However, in contrast with the naive expectation, 
the result for the total energy is not proportional to $x^2$, but rather to $x^{7/3}$. This is directly due to the Fermi exclusion
principle between polarons as we will see. Indeed, qualitatively, this makes it impossible for two polarons to be in the same state,
and depresses their probability to be spatially nearby, which decreases their overlap. 

Apparently this argument leads to the surprising conclusion that there is no contribution proportional to $x^2$.
However this is clearly incorrect. Indeed
such a contribution arises because, when we calculate the chemical potential of a single polaron and want to have our
result for the total energy correct up to order $x^2$, we have to take into account that the Fermi sea of the spin-up fermions is modified at first order in $x$ by the
presence of the spin-down fermions, which gives rise to a correction of order $x$ to the chemical potential of the polarons
in the dilute regime, and consequently a contribution of order $x^2$ for the overall energy.
Moreover, in order to obtain a proper expression of the polaron
binding energy, we have also to take into account the modification of the spin-down properties,
namely the existence of the polaron Fermi sea, which
is enough at the order we are working. One could expect that this last correction gives another contribution of order $x^2$
for the total energy, but it happens to be of order $x^{7/3}$, that is of the same order as the
term resulting from the overlap of the polarization clouds.

In the following we will calculate in turn these three above contributions. We will systematically stay at the simplest level
of approximation, namely restricting the spin-up Fermi sea excitations to single particle-hole excitations. This is first for
simplicity since the involved calculations are already not so simple. On the other hand this
is also justified by the fact that this approximation has proved to be excellent for the calculation of the polaron binding energy
and for its effective mass \cite{fred,crlc,ps1,ps2,nnj,nnj1}, for reasons which are quite well understood \cite{cg}. We may expect these same reasons to
carry over into our problem, although we will not attempt to prove it. There is no difficulty in principle to go to the level of two
particle-hole excitations or more within the framework we use, but in practice this would imply much more work. In the concluding part
we will come back to the discussion of the power expansion and also compare our results to recent works by Mora and Chevy \cite{chrisfred},
and by Yu, Z\"ollner and Pethick \cite{yzp} who have adressed basically the same problem as the one we consider here.

\section{Overlap of polarization clouds}\label{polar}
Our purpose in this section is to calculate the first correction to the dilute limit, corresponding to the term
with lowest power in $x$ related to the effect of the overlap between polarization clouds. 
We will proceed in the following way. 
We will first consider two polarons with momenta ${\bf p}_1$
and ${\bf p}_2$ and calculate their total energy. Their interaction energy ${\mathcal U}({\bf p}_1,{\bf p}_2)$ is the difference between 
this total energy and the sum of the energies
of the isolated single polarons with momenta ${\bf p}_1$ and ${\bf p}_2$ respectively. Then the contribution $E_{\rm int}^{(1)}$ to the total interaction energy coming from the overlap is obtained by
summing up over the free polaron Fermi sea, with radius $p_F$:
\begin{eqnarray}\label{intenerg}
E_{\rm int}^{(1)}=\frac{1}{2} \sum_{p_1,p_2 < p_F} {\mathcal U}({\bf p}_1,{\bf p}_2)
\end{eqnarray}
with the factor $1/2$ to avoid double-counting. Clearly this leaves out a number of effects which would lead to higher order corrections. 
For example the interaction energy of three polarons is not in general obtained by merely summing the three polaron-polaron
interactions, there are specific three-body contributions. Similarly we are allowed to consider the free polaron Fermi sea only
at lowest order, since in general the Fermi sea itself will be modified by interactions. However taking into account such an effect
would again lead to higher order terms in the expansion of the energy in powers of the density $x$.


In order to perform the above program we will extend our preceding work \cite{cg}, where we were dealing with a single spin-down interacting with a
free spin-up Fermi sea, to the case where we have two spin-down fermions. In principle we would like to write and solve 
the full corresponding Schr\"odinger equation. Actually this full problem is much too difficult and we will restrict ourselves  to a much simpler one
by restricting as much as possible the Hilbert space. In addition to the two spin-down fermions we allow only the presence
of a single particle-hole excitation in the spin-up Fermi sea. Restricting in such a way the Hilbert space is equivalent to perform
a variational calculation. Naturally the first justification for such a reduction of the Hilbert space is simplicity. However we believe that we will
nevertheless obtain a fairly good result for polaron interaction. Indeed it has been shown that, for the single polaron problem,
allowing a single particle-hole excitation provides an excellent result \cite{fred,crlc,cg} for the polaron chemical potential.
It is reasonable to believe that the reasons behind the success of this approximation \cite{cg} will also work for the case of two polarons.
However we will not try to prove it because we would need to allow at least two particle-hole excitations which, although quite possible,
is much more complicated than the calculation we will perform here.

Accordingly we consider for the two polarons an eigenstate of the form:
\begin{eqnarray}\label{wavef}
|\psi\rangle&=& \sum_{{\bf p}_1{\bf p}_2}\alpha _{{\bf p}_1{\bf p}_2} b^{\dag}_{{\bf p}_1}b^{\dag}_{{\bf p}_2}|0\rangle+
\sum_{{{\bf p}_1}{{\bf p}_2{\bf k}{\bf q}}}\alpha _{{\bf p}_1{\bf p}_2{\bf k}{\bf q}} \,
b^{\dag}_{{\bf p}_1+{\bf q}-{\bf k}}b^{\dag}_{{\bf p}_2}c^{\dag}_{\bf k}c_{\bf q}|0\rangle
\end{eqnarray}
where $|0\rangle=\prod_{k<k_F} c^{\dag}_{\bf k}\,|vac\rangle$ is the Fermi sea of $\uparrow$-spins,
$c_{{\bf k}}$ and $c^{\dag}_{{\bf k}}$ are
annihilation and creation operators for $\uparrow$-spin atoms while $b_{{\bf p}}$ and $b^{\dag}_{\bf p}$ are the corresponding operators
for the $\downarrow$-spin atoms. 
In the following we assume implicitly $k>k_F$ and $q<k_F$. Since the operators $b^{\dag}_{{\bf p}_1}$ and $b^{\dag}_{{\bf p}_2}$ anticommute, the coefficient
$\alpha _{{\bf p}_1{\bf p}_2}$ is naturally antisymmetric with respect to the exchange of ${\bf p}_1$ and ${\bf p}_2$:
\begin{eqnarray}\label{antialpp}
\alpha _{{\bf p}_1{\bf p}_2}=-\alpha _{{\bf p}_2{\bf p}_1}
\end{eqnarray}
implying the Pauli exclusion principle.
Similarly in the second term, by anticommuting the two $b^{\dag}$ operators and making the appropriate change of variable,
we conclude that:
\begin{eqnarray}\label{antialppkq}
\alpha _{{\bf p}_1{\bf p}_2{\bf k}{\bf q}}=-\alpha _{{\bf p}_2+{\bf k}-{\bf q},{\bf p}_1+{\bf q}-{\bf k},{\bf k}{\bf q}}
\end{eqnarray}
Hence we have for example in the first term of Eq.(\ref{wavef}) two identical terms describing a spin-down fermion with wavevector ${\bf p}_1$
together with a spin-down fermion with wavevector ${\bf p}_2$. To correct for this redundancy we should put a factor $1/2$ in front of the
eigenstate Eq.(\ref{wavef}). However since the normalisation of the wavefunction is unimportant in the following, it is simpler not to write this
corrective coefficient.

The Hamiltonian of our problem is the standard one channel Hamiltonian \cite{gps}, widely used to investigate the single polaron problem, namely:
\begin{eqnarray}\label{hamilt}
H&=&H_c+V \\ \nonumber
H_c&=& \sum_{\bf p}E({\bf p})b^{\dag}_{{\bf p}}b_{\bf p}+ \sum_{\bf k}\epsilon _{\bf k}c^{\dag}_{\bf k}c_{\bf k}\\  \nonumber
V&=&g \sum_{\bf k k' p p'}\delta_{\bf k k' p p'}c^{\dag}_{\bf k}c_{\bf k'}b^{\dag}_{{\bf p}}b_{\bf p'}
\end{eqnarray}
where $\epsilon_{\bf k} ={\bf k}^{2}/2m_{\uparrow}$ and $\,E({\bf p})={\bf p}^2/2m\ds$ are the kinetic energies (we take $\hbar=1$ throughout
the paper). In the interaction energy term $V$, the Kronecker symbol insures momentum conservation. Explicitly $\delta_{\bf k k' p p'}
\equiv \delta_{{\bf k}+{\bf p}-{\bf k}'-{\bf p}'}$.

In contrast with the single polaron case, where it is convenient to set the volume equal to unity, it is here necessary to write
explicitly the volume ${\mathcal V}$ in all the formulae. Indeed when we let this volume go to infinity, we will have physically
two independent polarons with a total energy equal to the sum of the energies of each polaron. If they have ${\bf p}_1$ and
${\bf p}_2$ as respective momenta, their energies are respectively $\mu + p^{2}_{1}/2m^{*}$ and $\mu + p^{2}_{2}/2m^{*}$ where
$\mu \equiv \mu \ds$ and $m^{*}$ are the chemical potential and the effective mass of a single polaron \cite{fred,crlc,ps1,ps2,cg}, which depend
naturally on the masses of the atoms, on the density $n=k^{3}_{F}/6\pi^2$ of the spin-up atoms and on the scattering length $a$.
On the other hand if the volume ${\mathcal V}$ is large but finite we expect, in addition to the preceding contributions to the total
energy, an interaction energy which will scale as $1/(k^{3}_{F}{\mathcal V})$, since the typical volume of the polarization cloud of spin-up
atoms surrounding a spin-down one is of order $1/k^{3}_{F}$, which has to be compared to the sample volume ${\mathcal V}$.
Hence we have to consider a finite volume if we want, for the interaction energy we are interested in, to recover a nonzero result.

This can be done in two equivalent procedures. A standard one is to introduce a factor $1/{\mathcal V}$ in the right-hand side of the
interaction term in Eq.(\ref{hamilt}). In this case the dimension of $g$ changes and, instead of being an energy, it becomes an energy 
multiplied by a volume. A completely equivalent, and for our purpose slightly more convenient, way is to keep Eq.(\ref{hamilt})
unchanged so that $g$ has still the dimension of an energy. In this case the relation between the scattering length $a$ and the interaction $g$ has to be modified.
For a unit volume this relation reads $m_r/(2\pi a)=g^{-1}+ \sum_{}^{k_c}2m_r/ k^2$, where $m_r = m\us m\ds /(m\us + m\ds)$  is the reduced
mass and $k_c$ is an upper cut-off in the interaction energy necessary to avoid ultraviolet divergences. As usual we will let this cut-off go to infinity 
while the coupling constant $g$ goes to zero, keeping the scattering length finite in the preceding relation. Physically this corresponds to take the
limit of a very short ranged interaction potential. However in the general case the dimensions of the preceding formula are incorrect and we have to write instead:
\begin{eqnarray}\label{regular}
\frac{{\mathcal V}\,m_r}{2\pi a}=\frac{1}{g}+ \sum_{0}^{k_c} \frac{2m_r}{k^2}
\end{eqnarray}
where one sees that all the terms have the same dimension. This equation is clearly equivalent to the equation obtained by the first procedure
(which is merely obtained by dividing by ${\mathcal V}$ and making the appropriate change of notation).
It may also be checked in the situation where $m\ds=\infty$, in which case we have merely a scattering potential located at the center of
a box of volume ${\mathcal V}$. This equation is convenient because nothing is changed compared to the case where the volume is unity,
except that we have to make the single simple replacement $a \rightarrow a/{\mathcal V}$.

\subsection{Projected Schr\"odinger equation}

Now, we write the Schr\"odinger equation $H|\psi\rangle=E|\psi\rangle$ and project it on the subspace corresponding to Eq.(\ref{wavef}),
i.e. onto the full Fermi sea and the Fermi sea with in addition a single particle-hole pair.
This yields a set of two coupled equations. We will not write them for the general case, but rather take advantage of simplifications which 
were already arising in the case of a single polaron \cite{cg}. First some terms, like the Hartree term, disappear in the limit of vanishing interaction
strength $g \rightarrow 0$ which we have to take, as we have explained above. Naturally we will not write these terms. Second the terms
resulting from the scattering of an spin-up particle (meaning with wavevector larger than $k_F$) display ultraviolet divergences analogous to
the one present in the last term in the right-hand side of Eq.(\ref{regular}), linked to the fact that the wavevector can go to infinity. 
These will be precisely cured by making use of Eq.(\ref{regular}). On the other hand terms physically related to the scattering of holes
(which have wavevectors smaller than $k_F$) do naturally not present these ultraviolet divergences. Hence in the above limit of 
$k_c \rightarrow \infty, g\rightarrow 0$ they are negligible compared to the preceding ones. Hence we will take only into account terms corresponding to
the scattering of spin-up particles and will not write those corresponding to the scattering of holes, because in the end these last ones turn out indeed
to be vanishing as can be checked by keeping them all the way long. With these simplifications we end up with:
\begin{eqnarray}\label{eqapp}
-g^{-1}\alpha _{{\bf p}_1{\bf p}_2}\left[E({\bf p}_1)+E({\bf p}_2)-E)\right]= \sum_{{\bf k}{\bf q}}\left(\alpha _{{\bf p}_1{\bf p}_2{\bf k}{\bf q}}
-\alpha _{{\bf p}_2{\bf p}_1{\bf k}{\bf q}}\right)
\end{eqnarray}
and
\begin{eqnarray}\label{eqappkq}
-g^{-1}\alpha _{{\bf p}_1{\bf p}_2{\bf k}{\bf q}}\,E^{(1)}_{{\bf p}_1{\bf p}_2{\bf k}{\bf q}}=\alpha _{{\bf p}_1{\bf p}_2}-
\alpha _{{\bf p}_2+{\bf k}-{\bf q},{\bf p}_1+{\bf q}-{\bf k}}+
 \sum_{{\bf K}}\left(\alpha _{{\bf p}_1{\bf p}_2{\bf K}{\bf q}}-\alpha _{{\bf p}_2+{\bf k}-{\bf q},{\bf p}_1+{\bf q}-{\bf k},{\bf K}{\bf q}}\right)
\end{eqnarray}
where we have set:
\begin{eqnarray}\label{defE1}
E^{(1)}_{{\bf p}_1{\bf p}_2{\bf k}{\bf q}}=\epsilon _{{\bf k}}-\epsilon _{{\bf q}}+E({\bf p}_1+{\bf q}-{\bf k})+E({\bf p}_2)-E
\end{eqnarray}
(we expect naturally $E<0$). We note that the antisymmetry properties Eq.(\ref{antialpp}) and Eq.(\ref{antialppkq}) are automatically 
satisfied by these equations. This has been obtained by making use, when necessary, of the anticommutation property of the two $b^{\dag}$ operators.

Then we manipulate and simplify these equations in much the same way as we have done in the case of a single polaron \cite{cg}.
Indeed the complication present here is the dependence of the wavefunction on the spin-down variables ${\bf p}_1$ and ${\bf p}_2$.
However the key part of the manipulation comes from the dependence on ${\bf k}$, in particular for large $k$, and this aspect is
essentially unchanged. Indeed it is clear from Eq.(\ref{eqapp}) that $\sum_{\bf k}\alpha _{{\bf p}_1{\bf p}_2{\bf k}{\bf q}} \sim g^{-1}$
since the summation over the hole variable ${\bf q}$ plays an unimportant role.
Dividing Eq.(\ref{eqappkq}) by $E^{(1)}_{{\bf p}_1{\bf p}_2{\bf k}{\bf q}}$ and summing over ${\bf k}$
leads to:
\begin{eqnarray}\label{intermed}
-g^{-1}\sum_{\bf k}\alpha _{{\bf p}_1{\bf p}_2{\bf k}{\bf q}}=\left(\alpha _{{\bf p}_1{\bf p}_2}+ \sum_{{\bf K}}\alpha _{{\bf p}_1{\bf p}_2{\bf K}{\bf q}}\right)
\left( \sum_{\bf k}\frac{1}{E^{(1)}_{{\bf p}_1{\bf p}_2{\bf k}{\bf q}}}
+\frac{{\mathcal V}\,m_r}{2\pi a}-g^{-1}- \sum_{0}^{k_c} \frac{2m_r}{k^2}\right) \\ \nonumber
- \sum_{\bf k}\frac{1}{E^{(1)}_{{\bf p}_1{\bf p}_2{\bf k}{\bf q}}}\left(\alpha _{{\bf p}_2+{\bf k}-{\bf q},{\bf p}_1+{\bf q}-{\bf k}}+
 \sum_{{\bf K}}\alpha _{{\bf p}_2+{\bf k}-{\bf q},{\bf p}_1+{\bf q}-{\bf k},{\bf K}{\bf q}}\right)
\end{eqnarray}
where we have introduced Eq.(\ref{regular}). The dominant contribution $g^{-1}\sum_{\bf k}\alpha _{{\bf p}_1{\bf p}_2{\bf k}{\bf q}}$ cancels out.
Then in the last parenthesis the first term $\alpha _{{\bf p}_2+{\bf k}-{\bf q},{\bf p}_1+{\bf q}-{\bf k}}$ is negligible compared to the last one
$ \sum_{{\bf K}}\alpha _{{\bf p}_2+{\bf k}-{\bf q},{\bf p}_1+{\bf q}-{\bf k},{\bf K}{\bf q}} \sim g^{-1}$ in the limit $g\rightarrow 0$.
We are left with:
\begin{eqnarray}\label{intermed1}
\left( \sum_{\bf k}\frac{1}{E^{(1)}_{{\bf p}_1{\bf p}_2{\bf k}{\bf q}}}
+\frac{{\mathcal V}\,m_r}{2\pi a}- \sum_{0}^{k_c} \frac{2m_r}{k^2}\right)\sum_{{\bf K}}\alpha _{{\bf p}_1{\bf p}_2{\bf K}{\bf q}} \\ \nonumber
=g^{-1}\alpha _{{\bf p}_1{\bf p}_2}+\sum_{\bf k}\frac{1}{E^{(1)}_{{\bf p}_1{\bf p}_2{\bf k}{\bf q}}}
\sum_{{\bf K}}\alpha _{{\bf p}_2+{\bf k}-{\bf q},{\bf p}_1+{\bf q}-{\bf k},{\bf K}{\bf q}}\
\end{eqnarray}
where in the factor of $\alpha _{{\bf p}_1{\bf p}_2}$, in the first term of the right-hand side, we have kept only the dominant term $-g^{-1}$,
the other ones being negligible in the limit $g\rightarrow 0$. Setting:
\begin{eqnarray}\label{defF}
F_{\bf q}({\bf p}_1,{\bf p}_2) \equiv g \sum_{{\bf K}}\alpha _{{\bf p}_1{\bf p}_2{\bf K}{\bf q}}
\end{eqnarray}
which is finite in the limit $g\rightarrow 0$, and with the notation:
\begin{eqnarray}\label{defT}
\left[T_{\bf q}({\bf p}_1,{\bf p}_2)\right]^{-1} \equiv \sum_{\bf k}\frac{1}{E^{(1)}_{{\bf p}_1{\bf p}_2{\bf k}{\bf q}}}
+\frac{{\mathcal V}\,m_r}{2\pi a}- \sum_{0}^{k_c} \frac{2m_r}{k^2}
\end{eqnarray}
we end up with:
\begin{eqnarray}\label{eqF}
\alpha _{{\bf p}_1{\bf p}_2}=\frac{F_{\bf q}({\bf p}_1,{\bf p}_2)}{T_{\bf q}({\bf p}_1,{\bf p}_2)} -  \sum_{k>k_F}
\frac{F_{\bf q}({\bf p}_2+{\bf k}-{\bf q},{\bf p}_1+{\bf q}-{\bf k})}{E^{(1)}_{{\bf p}_1{\bf p}_2{\bf k}{\bf q}}}
\end{eqnarray}
If we look at the volume dependence of the various terms, we see from Eq.(\ref{defT}), after making the conversion from summation over ${\bf k}$
to integration over ${\bf k}$ by $\sum_{\bf k} \rightarrow {\mathcal V}/(2\pi )^3 \int\,d{\bf k}$, that $\left[T_{\bf q}({\bf p}_1,{\bf p}_2)\right]^{-1}$
is proportional to volume ${\mathcal V}$.

It is convenient to remark that this equation relates terms in the wave function corresponding to a fixed value of ${\bf P}={\bf p}_1+{\bf p}_2$,
which is natural since the total momentum is conserved in the scattering of the two spin-down particles. Setting ${\bf p}_1={\bf P}/2+{\bf s}$
and ${\bf p}_2={\bf P}/2-{\bf s}$, with ${\bar \alpha} _{\bf P}({\bf s}) \equiv \alpha _{{\bf p}_1{\bf p}_2}$ and ${\bar F}_{\bf q P}({\bf s})
\equiv F_{\bf q}({\bf p}_1,{\bf p}_2)$, and similar notations for $T$ and $E$, we can rewrite Eq.(\ref{eqF}):
\begin{eqnarray}\label{eqF1}
{\bar \alpha} _{\bf P}({\bf s})=\frac{{\bar F}_{\bf q P}({\bf s})}{{\bar T}_{\bf q P}({\bf s})} -  \sum_{k>k_F}
\frac{{\bar F}_{\bf q P}({\bf k}-{\bf s}-{\bf q})}{{\bar E}^{(1)}_{{\bf P}{\bf s}{\bf k}{\bf q}}}
\end{eqnarray}

For fixed ${\bf P}$ and ${\bf q}$ this is just a matrix relation between the vectors ${\bar \alpha} _{\bf P}({\bf s})$ and ${\bar F}_{\bf q P}({\bf s})$.
Making the change ${\bf k}={\bf s}+{\bf t}+{\bf q}$, we can write:
\begin{eqnarray}\label{eqF2}
{\bar \alpha} _{\bf P}({\bf s})={\mathcal V}\, \sum_{\bf t}A_{{\bf q}{\bf P}}({\bf s},{\bf t}){\bar F}_{\bf q P}({\bf t})
\end{eqnarray}
where the matrix elements of $A$ are given by:
\begin{eqnarray}\label{elA}
A_{{\bf q}{\bf P}}({\bf s},{\bf t})=\frac{\delta_{{\bf s},{\bf t}}}{{\bar t}_{\bf q P}({\bf s})}
-\frac{1}{{\mathcal V}}\frac{\theta(|{\bf s}+{\bf t}+{\bf q}|-k_F)}{{\bar E}^{(1)}_{{\bf P}{\bf s},{\bf s}+{\bf t}+{\bf q},{\bf q}}}
\end{eqnarray}
and we have set:
\begin{eqnarray}\label{Ttbar}
{\bar T}_{\bf q P}({\bf s})=\frac{1}{{\mathcal V}}\,{\bar t}_{\bf q P}({\bf s})
\end{eqnarray}
Since:
\begin{eqnarray}\label{defE2}
{\bar E}^{(1)}_{{\bf P}{\bf s},{\bf s}+{\bf t}+{\bf q},{\bf q}}=\epsilon _{{\bf s}+{\bf t}+{\bf q}}-\epsilon _{{\bf q}}
+E(\frac{\bf P}{2}-{\bf s})+E(\frac{\bf P}{2}-{\bf t})-E={\bar E}^{(1)}_{{\bf P}{\bf t},{\bf s}+{\bf t}+{\bf q},{\bf q}}
\end{eqnarray}
the $A$ matrix is symmetric:
\begin{eqnarray}\label{Asym}
A_{{\bf q}{\bf P}}({\bf s},{\bf t})=A_{{\bf q}{\bf P}}({\bf t},{\bf s})
\end{eqnarray}
With our new notations Eq.(\ref{eqapp}) reads:
\begin{eqnarray}\label{eqapp1}
E\,{\bar \alpha} _{\bf P}({\bf s})= \left[\frac{1}{2}E({\bf P})+2E({\bf s})\right]{\bar \alpha} _{\bf P}({\bf s})
+ \sum_{{\bf q}}\left({\bar F}_{\bf q P}({\bf s})
-{\bar F}_{\bf q P}({\bf -s})\right)
\end{eqnarray}
where the two first terms $(1/2)E({\bf P})+2E({\bf s})=E({\bf p}_1)+E({\bf p}_2)$ represent the kinetic energy of the two spin-down particles
written as the sum of the kinetic energy associated to the center of mass motion and the kinetic energy of the relative motion.
Inverting Eq.(\ref{eqF2}) and carrying it into Eq.(\ref{eqapp1}) we obtain:
\begin{eqnarray}\label{eqapp2}
 \left[E\,- \frac{1}{2}E({\bf P})-2E({\bf s})\right]{\bar \alpha} _{\bf P}({\bf s})
&=& \frac{1}{{\mathcal V}}\,\sum_{{\bf q t}}\left(A^{-1}_{{\bf q}{\bf P}}({\bf s},{\bf t})-A^{-1}_{{\bf q}{\bf P}}({\bf -s},{\bf t})\right) {\bar \alpha} _{\bf P}({\bf t}) \\ \nonumber
&=& \frac{1}{{\mathcal V}}\,\sum_{{\bf q t}}\left(A^{-1}_{{\bf q}{\bf P}}({\bf s},{\bf t})+A^{-1}_{{\bf q}{\bf P}}({\bf -s},{\bf -t})\right) {\bar \alpha} _{\bf P}({\bf t})
\end{eqnarray}
where we have used Eq.(\ref{antialpp}) in the last step.
We will now make use of the explicit expression Eq.(\ref{elA}) for $A_{{\bf q}{\bf P}}({\bf s},{\bf t})$. 

\subsection{Infinite volume}\label{infvol}

We consider first the infinite volume limit where
the second term in the right-hand side of Eq.(\ref{elA}) is zero. In this case $A$ is diagonal and its inverse is:
\begin{eqnarray}\label{invAdiag}
A^{-1}_{{\bf q}{\bf P}}({\bf s},{\bf t})={\bar t}_{\bf q P}({\bf s})\,\delta_{{\bf s},{\bf t}}
\end{eqnarray}
Carrying this expression into Eq.(\ref{eqapp2}) and going to integration over ${\bf q}$ we have:
\begin{eqnarray}\label{eqapp3}
 \left[E\,- \frac{1}{2}E({\bf P})-2E({\bf s})\right]{\bar \alpha} _{\bf P}({\bf s})
&=& \frac{1}{{\mathcal (2\pi )^3}}\, \int d{\bf q}\,\left({\bar t}_{\bf q P}({\bf s})+{\bar t}_{\bf q P}({\bf -s})\right) {\bar \alpha} _{\bf P}({\bf s})
\end{eqnarray}
or with our original notations (with $t_{\bf q}({\bf p}_1,{\bf p}_2)={\bar t}_{\bf q P}({\bf s})$):
\begin{eqnarray}\label{eqapp4}
 \left[E\,-E({\bf p}_1)-E({\bf p}_2)\right]\, \alpha _{{\bf p}_1{\bf p}_2}
 &=& \frac{1}{{\mathcal (2\pi )^3}}\, \int d{\bf q}\,\left(t_{\bf q}({\bf p}_1,{\bf p}_2)+t_{\bf q}({\bf p}_2,{\bf p}_1)\right)\,\alpha _{{\bf p}_1{\bf p}_2}
\end{eqnarray}

The volume has disappeared from the equation and at first this result looks like what is expected for the energy of two non-interacting polarons.
Indeed let us consider the first term in the right-hand side. It gives the chemical potential of the first polaron ${\bf p}_1$. Indeed for this polaron
the second polaron ${\bf p}_2$ is irrelevant, and the relevant energy for this first polaron is $E-E({\bf p}_2)$, that is the total energy $E$ from which
the kinetic energy $E({\bf p}_2)$ of the second polaron has been removed. Similarly one can check that this is also this difference $E-E({\bf p}_2)$
which is entering $t_{\bf q}({\bf p}_1,{\bf p}_2)$ through Eq.(\ref{defE1}) and Eq.(\ref{defT}). With the change in notations $E-E({\bf p}_2) \rightarrow E$,
one recovers for this polaron:
\begin{eqnarray}\label{eqapp5}
E=E({\bf p}_1)
 &+& \int d{\bf q}\,\left[
   \int d{\bf k}\,\frac{1}{\epsilon _{{\bf k}}-\epsilon _{{\bf q}}+E({\bf p}_1+{\bf q}-{\bf k})-E}
+\frac{(2\pi )^2 m_r}{a}-  \int _{0}^{k_c}\!d{\bf k}\,\frac{2m_r}{k^2}\right]^{-1}
\end{eqnarray}
which is exactly \cite{fred,crlc,cg} the implicit equation for $E$ which gives the single polaron chemical potential together with its effective mass. 
Finally we can see in the same way
the second term in the right-hand side of Eq.(\ref{eqapp4}) as giving the chemical potential of the isolated second polaron ${\bf p}_2$ since, by
exchanging the variables ${\bf p}_1$ and ${\bf p}_2$ and taking the antisymmetry Eq.(\ref{antialpp}) of $\alpha _{{\bf p}_1{\bf p}_2}$ into account, 
it becomes the first term of the equation and we can repeat the above analysis.

Actually the above argumentation is clearly not completely satisfactory since the analysis of the first polaron energy implies that $E=\mu +p^{2}_{1}/2m\ds^*
+p^{2}_{2}/2m\ds$ while the second one corresponds rather to $E=\mu +p^{2}_{2}/2m\ds^*+p^{2}_{1}/2m\ds$. Since in general $|{\bf p}_1| \neq |{\bf p}_2|$,
these two values are inconsistent. We might perhaps argue that, since we are anyway interested in small values of $|{\bf p}_1|$ and $ |{\bf p}_2|$, 
we should neglect kinetic energies and only the zeroth order result $E=\mu $ is relevant. However even in this case we see that the sum of the two
contributions in the right-hand side of Eq.(\ref{eqapp4}) gives $2\mu $ instead of $\mu $, so that this equation is actually not satisfied.

The reasons for these problems are physically quite clear. Since we have taken the very strong restriction Eq.(\ref{wavef}) for the Hilbert space,
we do not even allow the situation where each one of the polarons has its own single particle-hole dressing. Although it is not in general 
meaningful to attach the particle-hole pair to a specific spin-down particle, we may roughly say that when one spin-down particle is dressed
by the particle-hole pair, giving it the energy $\mu $ (for zero kinetic energy), the other spin-down particle is naked and its energy is just
its kinetic energy. In this perspective the expressions written in the above paragraph for the total energy $E$ become quite meaningful.

A tempting way to solve the preceding inconsistencies is to enlarge our Hilbert space by allowing the existence of two particle-hole pairs.
In this case we would have the possibility to have each spin-down particle dressed by one particle-hole pair, allowing a much more satisfactory
physical description of the two polarons. However this would not solve all the consistency problems since we would have the possibility that
one down-particle is dressed by two particle-hole pairs while the other down-particle is naked. This would be a situation analogous to the one
we are faced with, although it would clearly be much more complicated since we should treat the single polaron problem at the level of two
particle-hole pairs. It is nevertheless possible to check that, if we allow each spin-down particle to have a single particle-hole pair
dressing, the above inconsistencies disappear. Indeed we can consider the case where the two spin-down particles belong to different
systems, each spin-down particle polarizing a different Fermi sea. Hence one has two non-interacting systems.
Physically this is not so different from the case of two polarons in an
infinite volume, since the overlap between the two polarons will also be zero in this last case. The state describing this situation is merely
the tensorial product of the two polarons (each one having its single particle-hole pair dressing). Hence there is the possibility to have two
particle-hole pairs. One can then write the corresponding Schr\"odinger equation, which decouples naturally into two Schr\"odinger equations,
describing the two independent systems. Hence one can follow how the existence of the additional term with two particle-hole pairs solves
the above inconsistencies and allows the decoupling of the equations, leading to the obvious result that the total energy is the sum of the
energies of each polaron, each one being obtained by solving Eq.(\ref{eqapp5}). Since writing all this explicitly does not make any problem,
but is fairly lengthy and burdensome, the reading being uninspiring, with a result which is completely obvious physically, 
we will not display the corresponding equations and stay at the level of this physical discussion. Our final conclusion is that, since at this
stage everything reduces to the calculation of the energy of a single polaron, the energy $E$ which comes in the expression of
$t_{\bf q}({\bf p}_1,{\bf p}_2)$, through Eq.(\ref{defE1}) and Eq.(\ref{defT}), is the single polaron energy, i.e. the polaron chemical potential 
$\mu $ when we work at zeroth order in kinetic energies.

As a final remark we note that all the problems discussed above are hidden in the case of the weak coupling limit, where $t_{\bf q}({\bf p}_1,{\bf p}_2)=
2\pi a/m_r$ and each term in the right-hand side of Eq.(\ref{eqapp4}) gives merely a (mean-field) contribution $2n\ds \pi a/m_r$.

\subsection{Polaron interaction}

Let us now consider the case of a finite volume ${\mathcal V}$. In this case the second term in Eq.(\ref{elA}) is no longer zero. However, because
of the factor $1/{\mathcal V}$, all the corresponding matrix elements are very small in the limit of a very large volume compared to the matrix
elements arising from the first term. This makes the inversion of the $A$ matrix very easy, all the more since the first term corresponds to a
diagonal matrix. One obtains:
\begin{eqnarray}\label{elA-1}
A^{-1}_{{\bf q}{\bf P}}({\bf s},{\bf t})={\bar t}_{\bf q P}({\bf s})\,\delta_{{\bf s},{\bf t}}
+\frac{1}{{\mathcal V}}\,{\bar t}_{\bf q P}({\bf s})\,\frac{\theta(|{\bf s}+{\bf t}+{\bf q}|-k_F)}{{\bar E}^{(1)}_{{\bf P}{\bf s},{\bf s}+{\bf t}+{\bf q},{\bf q}}}
\,{\bar t}_{\bf q P}({\bf t}) \equiv {\bar t}_{\bf q P}({\bf s})\,\delta_{{\bf s},{\bf t}}+\frac{1}{{\mathcal V}}\,B_{{\bf q}{\bf P}}({\bf s},{\bf t})
\end{eqnarray}
where $\theta$ is the Heaviside function. Upon substitution in Eq.(\ref{eqapp2}) we find:
\begin{eqnarray}\label{eqapp6}
 \left[E\,- \frac{1}{2}E({\bf P})-2E({\bf s})\right]{\bar \alpha} _{\bf P}({\bf s})
&=& \frac{1}{{(2\pi )^3}}\, \int d{\bf q}\,\left[{\bar t}_{\bf q P}({\bf s})+{\bar t}_{\bf q P}({\bf -s})\right] {\bar \alpha} _{\bf P}({\bf s}) \\ \nonumber
&+&\frac{1}{{\mathcal V}}\, \sum_{\bf t} \frac{1}{{(2\pi )^3}}\, \int d{\bf q}\,\left[B_{{\bf q}{\bf P}}({\bf s},{\bf t})+
B_{{\bf q}{\bf P}}({\bf -s},{\bf -t})\right] {\bar \alpha} _{\bf P}({\bf t})
\end{eqnarray}
Making use again of ${\bar \alpha} _{\bf P}(-{\bf t})=-{\bar \alpha} _{\bf P}({\bf t})$ the last term in the right-hand side may also be rewritten as:
\begin{eqnarray}\label{intB}
\frac{1}{2{\mathcal V}}\, \sum_{\bf t} \frac{1}{{(2\pi )^3}}\, \int d{\bf q}\,\left[B_{{\bf q}{\bf P}}({\bf s},{\bf t})-B_{{\bf q}{\bf P}}({\bf s},-{\bf t})
-B_{{\bf q}{\bf P}}(-{\bf s},{\bf t})+B_{{\bf q}{\bf P}}({\bf -s},{\bf -t})\right] {\bar \alpha} _{\bf P}({\bf t})
\end{eqnarray}

Let us now consider what would happen if we had two spin-down particles interacting directly through a general non-local
interaction $V$ and being ruled accordingly by the Hamiltonian:
\begin{eqnarray}\label{haminter}
H= H_c+H_{int}= \sum_{\bf p}E({\bf p})b^{\dag}_{{\bf p}}b_{\bf p}+\frac{1}{2{\mathcal V}}\, \sum_{\bf K K' Q}V_{\bf Q}({\bf K},{\bf K}')
b^{\dag}_{{\bf K}+{\bf Q}/2}b^{\dag}_{-{\bf K}+{\bf Q}/2}b_{-{\bf K}'+{\bf Q}/2}b_{{\bf K}'+{\bf Q}/2}
\end{eqnarray} 
where, because hermiticity and antisymmetry under fermion exchange, we must have:
\begin{eqnarray}\label{condV}
V_{\bf Q}({\bf K},{\bf K}')=V^{*}_{\bf Q}({\bf K}',{\bf K})=-V_{\bf Q}(-{\bf K},{\bf K}')=-V_{\bf Q}({\bf K},-{\bf K}')
\end{eqnarray}
When one writes the corresponding Schr\"odinger equation, one finds instead of the last term of Eq.(\ref{eqapp6}):
\begin{eqnarray}\label{schrV}
\frac{1}{2{\mathcal V}}\, \sum_{\bf t} \left[V_{{\bf P}}({\bf s},{\bf t})-V_{{\bf P}}({\bf s},-{\bf t})\right] {\bar \alpha} _{\bf P}({\bf t})=
\frac{1}{{\mathcal V}}\, \sum_{\bf t} V_{{\bf P}}({\bf s},{\bf t})\,{\bar \alpha} _{\bf P}({\bf t})
\end{eqnarray}
where we have made use of the antisymmetry relations Eq.(\ref{condV}).

When we compare Eq.(\ref{schrV}) with Eq.(\ref{intB}), we see that we can interpret Eq.(\ref{eqapp6}) as the Schr\"odinger equation
for the two spin-down particles interacting through an effective potential given by:
\begin{eqnarray}\label{Veff}
V_{{\bf P}}({\bf s},{\bf t})=\frac{1}{2}\frac{1}{{(2\pi )^3}}\, \int d{\bf q}\,\left[B_{{\bf q}{\bf P}}({\bf s},{\bf t})-B_{{\bf q}{\bf P}}({\bf s},-{\bf t})
-B_{{\bf q}{\bf P}}(-{\bf s},{\bf t})+B_{{\bf q}{\bf P}}({\bf -s},{\bf -t})\right] 
\end{eqnarray}
We note that this expression satisfies as it should the requirements Eq.(\ref{condV}). Naturally this effective interaction is due physically
to the fact that the two polarons have to share some part of their polarization clouds.

It is worth noting that our approach is quite general. Restricting ourselves to a single particle-hole pair leads to the explicit expression
Eq.(\ref{elA-1}) for $B$. However, if we extend our Hilbert space and consider any number of particle-hole pairs, Eq.(\ref{eqapp}) 
is still valid. And by eliminating
the wavefunction components corresponding to more than a single particle-hole pair, we will end up with an equation similar to Eq.(\ref{eqF2}).
The explicit expression of matrix $A$ will be more complex, but otherwise we can proceed in exactly the same way as we have done
above to identify the effective interaction, and consequently obtain the interaction energy as we will do just below. Hence extension
to taking for example into account two particle-hole pairs can proceed in this way, with the same framework.

If we calculate now the mean value of the interaction energy in Eq.(\ref{haminter}) for a state where the two spin-down particles occupy
plane waves ${\bf p}_1$ and ${\bf p}_2$, namely the state $b^{\dag}_{{\bf p}_1}b^{\dag}_{{\bf p}_2}|vac\rangle$, we obtain easily:
\begin{eqnarray}\label{interenerg}
\langle vac|b_{{\bf p}_2}b_{{\bf p}_1}\,H_{int}\,b^{\dag}_{{\bf p}_1}b^{\dag}_{{\bf p}_2}|vac\rangle=\frac{2}{{\mathcal V}} \,V_{{\bf P}}({\bf s},{\bf s})
\end{eqnarray}
with again the notations ${\bf P}={\bf p}_1+{\bf p}_2$ and ${\bf s}=({\bf p}_1-{\bf p}_2)/2$. Making use of Eq.(\ref{Veff}) this leads for the
two polarons to an interaction energy:
\begin{eqnarray}\label{}
{\mathcal U}({\bf p}_1,{\bf p}_2)=\frac{1}{{\mathcal V}}\frac{1}{{(2\pi )^3}}\, \int d{\bf q}\,\left[B_{{\bf q}{\bf P}}({\bf s},{\bf s})-B_{{\bf q}{\bf P}}({\bf s},-{\bf s})
-B_{{\bf q}{\bf P}}(-{\bf s},{\bf s})+B_{{\bf q}{\bf P}}({\bf -s},{\bf -s})\right] 
\end{eqnarray}
Making use of Eq.(\ref{intenerg}) we obtain finally:
\begin{eqnarray}\label{internerg1}
E_{\rm int}^{(1)}=\frac{1}{2} \sum_{p_1,p_2 < p_F}{\mathcal U}({\bf p}_1,{\bf p}_2)=\frac{1}{2{\mathcal V}} \sum_{p_1,p_2 < p_F}
\frac{1}{{(2\pi )^3}}\, \int d{\bf q}\,\left[B_{{\bf q}{\bf P}}({\bf s},{\bf s})-B_{{\bf q}{\bf P}}({\bf s},-{\bf s})
-B_{{\bf q}{\bf P}}(-{\bf s},{\bf s})+B_{{\bf q}{\bf P}}({\bf -s},{\bf -s})\right] 
\end{eqnarray}
We could as well calculate directly the average of the interaction energy $H_{int}$ given by Eq.(\ref{haminter}) over the spin-down Fermi sea
$|0\rangle=\prod_{p<p_F} b^{\dag}_{\bf p}\,|vac\rangle$. We would obtain a direct term and an exchange term, leading to:
\begin{eqnarray}\label{internerg2}
E_{\rm int}^{(1)}=\frac{1}{2{\mathcal V}}\, \sum_{p_1,p_2 < p_F} \left[V_{{\bf P}}({\bf s},{\bf s})-V_{{\bf P}}({\bf s},-{\bf s})\right]
=\frac{1}{{\mathcal V}}\, \sum_{p_1,p_2 < p_F} V_{{\bf P}}({\bf s},{\bf s})
\end{eqnarray}
which is identical to Eq.(\ref{internerg1}) from Eq.(\ref{Veff}).

\subsection{Total interaction energy}

Let us now proceed to evaluate explicitly this interaction energy. We note first that, by going from summation to integration through 
$\sum_{{\bf p}_i} \rightarrow {\mathcal V}/(2\pi )^3 \int\,d{\bf p}_i$, we find that $E_{\rm int}^{(1)}$ is as expected an extensive quantity,
proportional to volume. Having settled this point, we will take in the following ${\mathcal V}=1$ for simplicity.
Then we notice that the dependence of $E_{\rm int}^{(1)}$ on $n\ds$ will be faster than the standard dependence in $n\ds^2 \sim p_F^6$ 
expected for a standard interaction. This is due to the fact that  $V_{{\bf 0}}({\bf 0},{\bf 0})=0$, so the naive answer 
$E_{\rm int}^{(1)}=n\ds^2 V_{{\bf 0}}({\bf 0},{\bf 0})$ does not hold. This cancellation, seen explicitly in Eq.(\ref{Veff}), results automatically
from the antisymmetry properties Eq.(\ref{condV}) and so it comes directly from the fact that polarons are fermions, just as the spin-down
particles themselves.

We will naturally use the fact that $p_F$ is small compared to $k_F$, and for consistency we have to make the calculation to lowest
order in $p_F$. However, in view of the above mentioned cancellation, we have to do it carefully. Looking at the explicit expression
Eq.(\ref{elA-1}) of $B_{{\bf q}{\bf P}}({\bf s},{\bf t})$, we see that if we set ${\bf s}={\bf t}={\bf 0}$ in the Heaviside function we obtain
$\theta(q-k_F)$, so that $q$ must be larger than $k_F$. However by definition $q<k_F$. Hence the result will be zero due to
this Heaviside function. This means that, although $|{\bf s}|$,$|{\bf t}|<p_F$ we have to keep carefully nonzero ${\bf s}$ and ${\bf t}$ in this Heaviside function.
On the other hand we may perfectly set ${\bf s}={\bf t}={\bf 0}$ in the other factors since they are not sensitive to this approximation
and the result is nonzero. Similarly $|{\bf P}|<2p_F$ allows us to set ${\bf P}={\bf 0}$ since the resulting values of the different factors
are nonzero. A further simplification comes from the fact that $B_{{\bf q}{\bf P}}({\bf s},-{\bf s})=B_{{\bf q}{\bf P}}(-{\bf s},{\bf s})=0$
since both are proportional to $\theta(q-k_F)$ which leads to a zero result as explained above. Finally exchanging ${\bf p}_1$ and ${\bf p}_2$
changes ${\bf s}$ into $-{\bf s}$, hence the last term in Eq.(\ref{internerg1}) gives the same result as the first one. We are left with:
\begin{eqnarray}\label{Eint0}
E_{\rm int}^{(1)}=\frac{[{\bar t}_{\bf k_F 0}({\bf 0})]^2}{|\mu|}\, \frac{1}{(2\pi )^9}\,\int_{p_1<p_F}d{\bf p}_1\int_{p_2<p_F}{\bf p}_2
\int_{q<k_F}d{\bf q}\;\theta(|{\bf p}_1-{\bf p}_2+{\bf q}|-k_F)
\end{eqnarray}
where we have used ${\bar E}^{(1)}_{{\bf 0}{\bf 0},{\bf q},{\bf q}}=|\mu |$ from Eq.(\ref{defE1}), since we have explained at
the end of section \ref{infvol} that we have to take $E=\mu $. We have also replaced ${\bar t}_{\bf q 0}({\bf 0})$
by ${\bar t}_{\bf k_F 0}({\bf 0})$, because to zeroth order in $p_F$ we have $q=k_F$ from the Heaviside function.

\begin{figure}
\includegraphics[width=\linewidth]{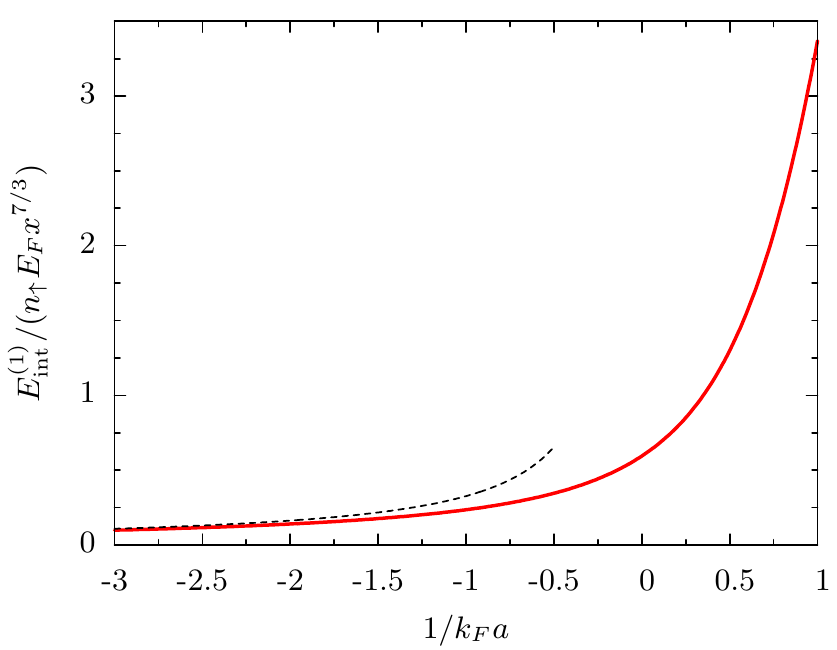}
\caption{\label{fig1} (Color online) $E_{\rm int}^{(1)}/(n\us E_F x^{7/3})$ as a function of
$1/k_F a$ in the case of equal masses $m\us=m\ds$. The dashed line is the corresponding expansion in the weak coupling limit.}
\end{figure}

Hence we are left to evaluate the high dimensional volume corresponding to the triple integration. The ${\bf q}$ integration is
easily performed. Indeed ${\bf q}$ must be inside a sphere of radius $k_F$ centered at the origin, but from the Heaviside
function outside a sphere of radius $k_F$ centered at $2{\bf s}={\bf p}_1-{\bf p}_2$. Thus the result is the volume $4\pi k^3_F/3$
of the whole sphere minus the volume of their intersection. Since the volume of the intersection of two spheres of radius $R$ with their center
distance being $r$ is given by $V_{\cap}=(\pi /3)(4R^3-3R^2 r+r^3/4)$, this gives:
\begin{eqnarray}\label{integrq}
\int_{q<k_F}d{\bf q}\;\theta(|{\bf p}_1-{\bf p}_2+{\bf q}|-k_F)=\frac{2\pi }{3}(3sk^2_F-s^3) \simeq 2\pi sk^2_F
\end{eqnarray}
where we have used $s \ll k_F$ in the last step. Setting ${\bf p}_1=p_F {\bf x}$ and ${\bf p}_2=-p_F {\bf y}$, we are left, apart
from a factor $p_F^7$, with the calculation of:
\begin{eqnarray}\label{intp1p2}
\int_{x<1}d{\bf x}\int_{y<1}d{\bf y}\;|{\bf x}+{\bf y}|
\end{eqnarray}
for which it is more convenient to take the new variables ${\bf u}={\bf x}+{\bf y}$ and ${\bf v}=({\bf x}-{\bf y})/2$, with $d{\bf x}\,d{\bf y}=
d{\bf u}\,d{\bf v}$. Since it is easy to see that, at fixed ${\bf u}$, vector ${\bf v}$ (with its origin at the middle of vector ${\bf u}$)
has to run inside the intersection of the two spheres of radius $1$, with distance $u$ between centers, the ${\bf v}$ integration gives
again $V_{\cap}$ with $R=1$ and $r=u$, i.e. $(\pi /3)(4-3u+u^3/4)$. Hence:
\begin{eqnarray}\label{intp1p2a}
\int_{x<1}d{\bf x}\int_{y<1}d{\bf y}\;|{\bf x}+{\bf y}|=\frac{\pi }{3}\, 4\pi  \int_{0}^{2}du\,u^3 (4-3u+\frac{u^3}{4})=\frac{64\pi ^2}{35}
\end{eqnarray}

This leads us finally to:
\begin{eqnarray}\label{EInt}
E_{\rm int}^{(1)}=\frac{[{\bar t}_{\bf k_F}]^2}{|\mu|}\, \frac{1}{(2\pi )^9}\frac{64\pi ^3}{35}\,p_F^7k_F^2=
\frac{[{\bar t}_{\bf k_F}]^2}{|\mu|}\, \frac{9}{35}\left(\frac{3}{4\pi ^4}\right)^{1/3}\,n\ds^{7/3}k_F^2
\end{eqnarray}
where ${\bar t}_{\bf k_F} \equiv {\bar t}_{\bf k_F 0}({\bf 0})$.
Hence we obtain an interaction energy which scales as $n\ds^{7/3}$, and not as $n\ds^2$ as discussed above. One can express
${\bar t}_{\bf k_F 0}({\bf 0})$ analytically in terms of $k_F$, $a$, $m\us$, $m\ds$ and $|\mu |$, but there is no point to write here
the somewhat complicated formula. In the weak coupling limit $a\rightarrow0_-$, it reduces to $2\pi a/m_r$. Since in this case
we have also the mean field value $|\mu |=2\pi n\us |a|/m_r$, we obtain explicitly:
\begin{eqnarray}\label{Eintwc}
E_{\rm int}^{(1)}=\frac{9}{35\pi }\,p_F |a| \frac{n\ds ^2}{n\us}\,\frac{k_F^2}{m_r}
\end{eqnarray}
Let us finally note that this interaction energy $E_{\rm int}^{(1)}$ is positive which corresponds to a net repulsion between two polarons.
This is already seen at the level of Eq.(\ref{Eint0}). This result is physically reasonable: two polarons have to compete to make up
their spin-up clouds, so that their dressing is not as optimal as the cloud of a fully isolated polaron.

In order to compare the above interaction energy to other contributions to the system total energy, 
it is convenient to write the dimensionless ratio $E_{\rm int}^{(1)}/n\us E_F$, where $E_F=\mu \us=k_F^2/2m \us$ and to introduce
the ratio $x=n\ds /n\us$. This leads to:
\begin{eqnarray}\label{EInt1}
\frac{E_{\rm int}^{(1)}}{n\us E_F}=\frac{3}{35\, \pi ^4}\,\frac{[m\us k_F {\bar t}_{\bf k_F}]^2}{\rho}\,x^{7/3}
\end{eqnarray}
where we have set $|\mu|=\rho E_F$. In the weak coupling limit, this result becomes 
$18 m\us k_F|a|\, x^{7/3} /(35 \pi m_r )$. For the case of equal masses $m\us=m\ds=2m_r$, the 
coefficient in front of $x^{7/3}$ in the result Eq.(\ref{EInt1}) is plotted in Fig.\ref{fig1}. We have
limited the plot range to $1/k_Fa < 1$, since we know \cite{ps2,gcl} that, for $1/k_Fa =0.88$,
the polaron becomes unstable with respect to the formation of a (dressed) bound state between
the spin-down and a spin-up fermion, so that beyond this point the whole physical picture breaks down.

\section{Corrections to the single polaron chemical potential}

Let us now come to the corrections we have to make for the single polaron chemical potential, due to the presence of
a non zero density of spin-down atoms. As we have already indicated in the introduction, they have two origins. The first one
is the modification to first order of the properties of the spin-up Fermi sea, which implies a corresponding change
of the spin-up propagator coming in the calculation. The second one is merely the existence of the polaron, or equivalently
spin-down Fermi sea. We begin by considering this last effect since it is by far the simpler one.

\subsection{Correction due to the polaron Fermi sea}\label{polarFerm}

When one calculates the chemical potential of a polaron, one assumes that, when a particle-hole pair is created, 
the spin-down atom may have any recoil to compensate for the momentum of this particle-hole pair. However in the presence
of a Fermi sea of spin-down particles this is no longer allowed since Pauli exclusion forbids all the momenta inside this Fermi
sea. Hence in Eq.(\ref{eqapp5}) which gives the chemical potential we are looking for (provided we set the momentum 
${\bf p}_1={\bf 0}$), we have to put in the sum over ${\bf k}$ a further restriction $|{\bf k}-{\bf q}|>p_F$. The resulting small
modification of the integral is:
\begin{eqnarray}\label{deltaI}
\delta I({\bf q})&=& \int d{\bf k}\,\frac{\theta(|{\bf k}-{\bf q}|-p_F)}{\epsilon _{{\bf k}}-\epsilon _{{\bf q}}+E({\bf q}-{\bf k})+|\mu |}
- \int d{\bf k}\,\frac{1}{\epsilon _{{\bf k}}-\epsilon _{{\bf q}}+E({\bf q}-{\bf k})+|\mu |} \\ \nonumber
&=&- \int d{\bf k}\,\frac{\theta(p_F-|{\bf k}-{\bf q}|)}{\epsilon _{{\bf k}}-\epsilon _{{\bf q}}+E({\bf q}-{\bf k})+|\mu |}
\end{eqnarray}
where we have naturally set $E=-|\mu |$ since we are looking for the lowest order correction. We have by definition
$|{\bf k}|>k_F$ and $|{\bf q}|<k_F$. Hence $\theta(p_F-|{\bf k}-{\bf q}|)$ implies that $|{\bf k}| \simeq |{\bf q}| \simeq k_F$
since $p_F$ is small. To lowest order the denominator is just equal to $|\mu |$ and we are left with the calculation
of $ \int \!d{\bf k}\,\theta(p_F-|{\bf k}-{\bf q}|)$ which is just the volume of the part of a small sphere of radius $p_F$,
centered at ${\bf q}$ and which is outside the sphere of radius $k_F$ centered at the origin. Since $p_F$ is small
we may consider the relevant part of this last sphere to be a plan, and we have just to find the volume of the $p_F$
sphere which is beyond a plan with nearest distance to the center of this sphere equal to $k_F-q$, with $q=|{\bf q}|$.
This volume is easily found to be $(\pi/3) [2p_F+k_F-q][p_F-(k_F-q)]^2$, with naturally the condition $0<k_F-q<p_F$,
leading to:
\begin{eqnarray}\label{deltaI1}
\delta I({\bf q})=-\frac{\pi }{3|\mu |}\,(2p_F+k_F-q)(p_F+q-k_F)^2\,\theta(p_F+q-k_F)
\end{eqnarray}
This modification $\delta I({\bf q})$ induces a shift $\delta|\mu |$ in the chemical potential. Differentiating Eq.(\ref{eqapp5})
with respect to $E$ we find:
\begin{eqnarray}\label{shift1}
\delta|\mu |=\frac{1}{Z}\,\frac{[{\bar t}_{\bf k_F}]^2}{(2\pi )^6}\, \int d{\bf q}\,\delta I({\bf q})
\end{eqnarray}
where we have used the fact that $\delta I({\bf q})$ is nonzero only for $q \simeq k_F$. We have set:
\begin{eqnarray}\label{defZ0}
Z=1+\frac{1}{(2\pi )^6} \int d{\bf q}\,[{\bar t}_{\bf q 0}({\bf 0})]^2 \, \int d{\bf k}
\frac{1}{\left[E^{(1)}_{{\bf 0}{\bf 0}{\bf k}{\bf q}}\right]^2}
\end{eqnarray}
in which the ${\bf k}$ integration can be performed analytically, but the ${\bf q}$ integration has to be done numerically. 
The integral in the right-hand side of Eq.(\ref{shift1})
is easily evaluated. This leads to:
\begin{eqnarray}\label{shift2}
\delta|\mu |=-\frac{1}{Z}\,\frac{[{\bar t}_{\bf k_F}]^2}{64\pi^4|\mu |} p_F^4 k_F^2 
\end{eqnarray}
The result is negative as expected, decreasing the binding of the spin-down in the spin-up Fermi sea, since the phase
space for creation of a particle-hole pair has been decreased. 

In order to obtain the corresponding contribution $E_{\rm int}^{(2)}$ 
to the interaction, we have to make use of the definition of the chemical
potential $\mu\ds =\partial \mathcal E /\partial n\ds$, where $\mathcal E$ is the total energy of the system.
Since we find that the above spin-down shift in chemical potential is proportional to $p_F^4 \sim n\ds^{4/3}$, 
the corresponding term in the interaction energy is:
\begin{eqnarray}\label{}
E_{\rm int}^{(2)}=-\frac{3}{7}\,n\ds \delta|\mu |
\end{eqnarray}
where $\delta|\mu |$ is given by Eq.(\ref{shift2}). This  contribution to the interaction energy is again a positive.

We note that this term is proportional to $p_F^7$. This is exactly the same dependence as the one we have found
for the interaction between polarons (and in the same way it increases the total energy). Hence this contribution should
not be overlooked. One could have naively expected a $p_F^3$ dependence for $\delta|\mu |$ in Eq.(\ref{shift2}),
corresponding to the volume of the polaron Fermi sea. The additional $p_F$ factor comes from the fact that only
a thin shell, of thickness $p_F$, is involved in the hole phase space, corresponding to the variable ${\bf q}$,
and not the complete hole phase space.

In the weak coupling limit $a\rightarrow0_-$ we have seen that ${\bar t}_{\bf k_F}=2\pi a/m_r$ and similarly
${\bar t}_{\bf q 0}({\bf 0})=2\pi a/m_r$, so the bracket in the right-hand side of Eq.(\ref{shift2}) reduces to unity
and we obtain for this contribution to the total energy:
\begin{eqnarray}\label{shiftwc}
E_{\rm int}^{(2)}=\frac{9}{112\pi  }\,p_F |a| \frac{n\ds ^2}{n\us}\,\frac{k_F^2}{m_r}
\end{eqnarray}
We see that it has exactly the same form as the interaction energy Eq.(\ref{Eintwc}), with a coefficient which is not
so different. Hence this effect can not be omitted.

\begin{figure}
\includegraphics[width=\linewidth]{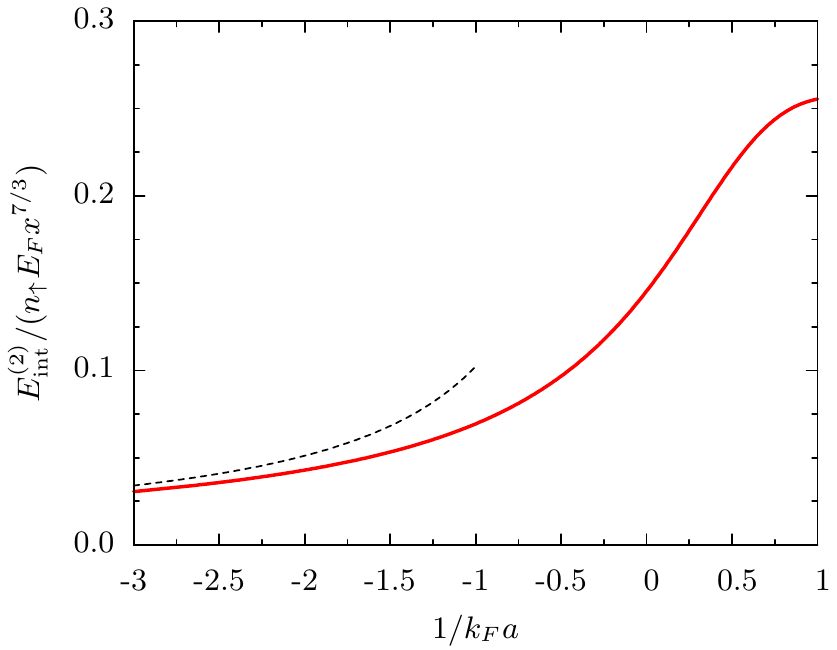}
\caption{\label{fig2} (Color online) $E_{\rm int}^{(2)}/(n\us E_F x^{7/3})$ as a function of
$1/k_F a$ in the case of equal masses $m\us=m\ds$. The dashed line is the corresponding expansion in the weak coupling limit.}
\end{figure}

Finally, from Eq.(\ref{shift2}),  the corresponding ratio $E_{\rm int}^{(2)} / n\us E_F$ is given by:
\begin{eqnarray}\label{EInt2}
\frac{E_{\rm int}^{(2)}}{n\us E_F}=\frac{3}{112\, \pi ^4}\,\frac{1}{Z}\,\frac{[m\us k_F {\bar t}_{\bf k_F}]^2}{\rho}\,x^{7/3}
\end{eqnarray}
The coefficient in front of $x^{7/3}$ in Eq.(\ref{EInt2}) is plotted in Fig.\ref{fig2}
for equal masses $m\us=m\ds$.

\subsection{Correction due to the modification of the spin-up Fermi sea}\label{polarFermup}

\begin{figure}
\includegraphics[width=\linewidth]{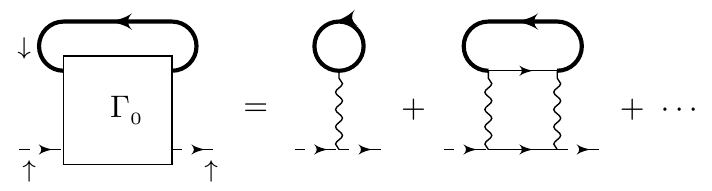}
\caption{\label{ladder} Ladder expansion for the spin-up self-energy Eq.(\ref{self}).}
\end{figure}

We consider now how the spin-up Fermi sea is modified by the spin-down population to first order in $n\ds$ and how
consequently the spin-down chemical potential $\mu \ds$ is changed to first order in $n\ds$.  
This leads to a contribution to the total energy proportional to $n\ds ^2$, that is exactly the
dependence expected from an interaction between spin-down particles.
The first stage is to find the spin-up self-energy. This is just a Hartree-like term coming from the spin-down population.
The only difference is that, since the $g$ goes to zero, we should not write merely a single interaction, but rather sum over
repeated interactions, which leads to sum up ladder diagrams (see Fig.\ref{ladder}). This gives:
\begin{eqnarray}\label{self}
\Sigma\us ({\bf k},\omega )= \sum_{\bf k'} \int  \frac{d\omega '}{2i\pi} G\ds({\bf k}', \omega ' )
\Gamma_0({\bf k}+{\bf k}',\omega +\omega ')
\end{eqnarray}
In agreement with our above notations we have used again the convention $ \sum_{\bf k} \equiv  \int d{\bf k}/(2\pi )^3$. The $\omega '$ integration runs over the imaginary frequency axis. Here $\Gamma_0({\bf k},\omega)$ is the vertex used already in Ref.\cite{crlc} to calculate actually the polaron chemical
potential (it contains the bare propagator $G_{0 \down}$).
Since the $G\ds$ factor in Eq.(\ref{self}) will give rise to a factor $n\ds$, we have naturally to evaluate the sum of the ladder
diagrams to zeroth order in $n\ds$, which is indicated by the subscript $0$. Explicitly:
\begin{eqnarray}\label{defG0}
[\Gamma_0(K,\Omega)]^{-1}=\frac{m_r}{2\pi a}- \sum_{\bf k}\left[\frac{2m_r}{k^2}
+\frac{\theta(\epsilon_k-\mu\us)}{\Omega+\mu\us+ \mu ^0\ds-\epsilon_k-E({\bf k}+{\bf K})}\right]
\end{eqnarray}
where the superscript $0$ indicates that the chemical potential has to be evaluated to zeroth order in $n\ds$.
Naturally $\Gamma_0(K,\Omega)$ is closely related to $T_{\bf q}({\bf p}_1,{\bf p}_2)$ defined above in Eq.(\ref{defT}).

Since we do not consider the possibility of a bound state between an up and a down particle, the only singularities of $\Gamma_0(K,\Omega)$
occur only on the positive real $\Omega $ axis (one can check that, for $\Omega <0$, the denominator in the integral in Eq.(\ref{defG0}) is always negative)
which correspond physically to the continuous spectrum of the scattering states of the up and a down particle. We calculate $\Sigma\us ({\bf k},\omega )$
from Eq.(\ref{self}) for imaginary frequency $\omega $ and then continue analytically the result for any frequency. In order to avoid the
singularities of $\Gamma_0$ we close the $\omega '$ path by a semi-circle at infinity in the ${\rm Re}\,\omega ' <0$ half-plane, so that the integration
contour encloses this half-plane. The only contribution comes from the pole of $G\ds({\bf k}', \omega ' )$. The down particle number $n\ds$
is given by:
\begin{eqnarray}\label{dnumb}
n\ds=\sum_{\bf k'} \int  \frac{d\omega '}{2i\pi} G\ds({\bf k}', \omega ' )
\end{eqnarray}
where the integration contour is the same as above, enclosing again the ${\rm Re}\,\omega ' <0$ half-plane. In the case of a single spin-down particle, where
$n\ds$ is vanishingly small, the ground state corresponds to a pole of $G\ds({\bf k}', \omega ' )$ at zero momentum ${\bf k}'=0$ and zero frequency $\omega '=0$. In the case
of a small number of spin-down particles, the poles will be similarly in the vicinity of ${\bf k}'=0$ and $\omega '=0$. Hence we can set 
${\bf k}'=0$ and $\omega '=0$ in $\Gamma_0({\bf k}+{\bf k}',\omega +\omega ')$ in Eq.(\ref{self}), so that we are left with an integral which is
just Eq.(\ref{dnumb}). This leads us to:
\begin{eqnarray}\label{self1}
\Sigma\us ({\bf k},\omega )=n\ds \Gamma_0({\bf k},\omega)
\end{eqnarray}
In particular in the weak coupling limit $a\rightarrow0_{-}$, where from Eq.(\ref{defG0}) $\Gamma_0(K,\Omega)=2\pi a/m_r$, we find:
\begin{eqnarray}\label{selfwc}
\Sigma\us ({\bf k},\omega )=\frac{2\pi a}{m_r}n\ds
\end{eqnarray}
which is just the expected mean-field result.

Having found the self-energy, we have the full spin-up propagator:
\begin{eqnarray}\label{Gup}
G\us({\bf k}, \omega )=\frac{1}{\omega -\epsilon_k+\mu \us- \Sigma\us ({\bf k},\omega )}\simeq
G_{0 \up}({\bf k}, \omega )+\Sigma \us ({\bf k},\omega )G^2_{0 \up}({\bf k}, \omega ) \equiv 
G_{0 \up}({\bf k}, \omega )+\delta G_{\up}({\bf k}, \omega )
\end{eqnarray}
where $G_{0 \up}({\bf k}, \omega )=[\omega -\epsilon_k+\mu \us]^{-1}$ is the free spin-up propagator.
Here we have in the last step expanded the result to first order in $n\ds$, consistent with our low spin-down density approximation.
With this result we may, in our grand-canonical framework (which is just the $T=0$ limit of the standard finite $T$ formalism), calculate
the change $\delta n\us$ of spin-up particle density resulting from the spin-down particle density $n\ds$, at fixed $\mu \us$.
It is given by:
\begin{eqnarray}\label{deltnup}
\delta n\us= \sum_{\bf k} \int  \frac{d\omega}{2i\pi} \delta G\us({\bf k}, \omega)=
\sum_{\bf k} \int  \frac{d\omega}{2i\pi} \Sigma \us ({\bf k},\omega )G^2_{0 \up}({\bf k}, \omega )=
-\frac{d}{dx}\sum_{\bf k} \int  \frac{d\omega}{2i\pi}\frac{ \Sigma \us ({\bf k},\omega )}{\omega -\epsilon_k+x} \;\Big|_{x=\mu \us}
\end{eqnarray}
Since, from Eq.(\ref{self1}), $\Sigma \us ({\bf k},\omega )$ is analytical for ${\rm Re}\,\omega <0$, it is convenient to close again the contour
around this negative frequency half-plane. The only contribution comes from the pole at $\omega =\epsilon_k-x$, which gives a factor
$\theta(x-\epsilon_k)$ since we want this pole to be at negative frequency. This leads to:
\begin{eqnarray}\label{deltnup1}
\delta n\us= \sum_{\bf k}\left[\theta(\mu \us-\epsilon_k) \frac{\partial \Sigma \us ({\bf k},\epsilon_k -\mu \us)}{\partial \omega }
-\delta(\epsilon_k -\mu \us) \Sigma \us ({\bf k},0)\right]
\end{eqnarray}
where $\partial \Sigma \us ({\bf k},\epsilon_k -\mu \us)/\partial \omega$ is a short-hand for the partial derivative $\partial \Sigma \us ({\bf k},\omega )/\partial \omega$
taken for $\omega =\epsilon_k -\mu \us$.

Interestingly, as pointed out in \cite{zbp}, $\delta n\us$ has a simple physical interpretation for $n\ds \to 0$. It is merely the number of spin-up particles in
the cloud forming a single polaron. This quantity, or equivalently $\nu=\partial n\us /\partial n\ds$ at fixed $\mu \us$ which is obtained from Eq.(\ref{deltnup1}) 
and Eq.(\ref{self1}), satisfies a simple thermodynamic identity \cite{zbp}. We have $\nu=- \partial \mu \ds /\partial \mu \us$, where the derivative is taken at
fixed $n\ds$. Taking $n\ds=0$ corresponds to the case of a single polaron and in this case in the simple framework used in Ref.\cite{crlc}, $\mu \ds$
is given by:
\begin{eqnarray}\label{formud0}
\mu^0 \ds= \sum_{\bf K} \theta(\mu \us-\epsilon_K) \Gamma_0({\bf K},\epsilon_K-\mu \us)
\end{eqnarray}
Taking the explicit derivative of this formula with respect to $\mu \us$ gives a result in full agreement with the thermodynamic identity.
However, since from Eq.(\ref{defG0}), $\Gamma_0({\bf K},\Omega )$ has also a dependence on $\mu \us$ (as well as on $\mu \ds$),
we should take also these dependences into account in the calculation of the derivative. As a result the thermodynamic identity is not
satisfied. This is not so surprising since Eq.(\ref{deltnup1}) and Eq.(\ref{formud0}) are only an approximate results. Nevertheless we have found numerically that
the disagreement with the thermodynamic identity stays almost always quite small, whatever the value of the parameter $1/k_Fa$. The
difference takes sizeable values only when one approaches the transition point $1/k_Fa \simeq 0.88$ where the polaron becomes unstable
with respect to the formation of a molecular bound state \cite{ps2,gcl}.

We consider now how the calculation of $\mu \ds$ is modified when we take into account that $G_{0 \up}$ has to be replaced by $G_{0 \up}+\delta G_{\up}$.
Taking again the framework used in Ref.\cite{crlc}, we have:
\begin{eqnarray}\label{formud}
\mu \ds=\Sigma\ds ({\bf 0},0 )= \sum_{\bf K} \int  \frac{d\Omega}{2i\pi} G\us({\bf K}, \Omega )
\Gamma({\bf K},\Omega)
\end{eqnarray}
Compared to the calculation in the single polaron limit $n\ds \rightarrow 0$ which leads to Eq.(\ref{formud0}), 
we have now to take into account that, in this formula, $G\us$ is no longer
$G_{0\up}$, but $G\us=G_{0\up}+\delta G_{\up}$ given by Eq.(\ref{Gup}). Similarly $\Gamma$ is no longer $\Gamma_0$, but $\Gamma=\Gamma_0+\delta \Gamma$,
since in summing up the ladder diagrams, we have now to take into account that the spin-up propagator $G\us$ is no longer $G_{0\up}$.
Since in this summation the contribution coming from the propagators between two interactions is basically a convolution
of the spin-up and spin-down propagators, the modification will come from the quantity:
\begin{eqnarray}\label{varGam}
 \sum_{\bf k} \int  \frac{d\omega}{2i\pi} G_{0\down}({\bf K}-{\bf k},\Omega- \omega) \delta G\us({\bf k}, \omega)=
 \sum_{\bf k} \int  \frac{d\omega}{2i\pi} G_{0\down}({\bf K}-{\bf k},\Omega- \omega)
\Sigma \us ({\bf k},\omega )G^2_{0 \up}({\bf k}, \omega )
\end{eqnarray}
where $G_{0\down}({\bf k}, \omega)=[\omega+ \mu ^0\ds-E({\bf k})]^{-1}$ is, consistently with our above framework, the down propagator
with the down self-energy set to zero, but with the proper polaron chemical potential. 
Since $\Sigma \us ({\bf k},\omega )$ is analytical for ${\rm Re}\,\omega <0$, it is again convenient to close the contour
around the negative frequency half-plane. Contributions will only come from the poles of $G_{0\down}$ and $G_{0 \up}$ located
in this domain. Using for the double pole due to $G_{0 \up}$ the same convenient trick as in Eq.(\ref{deltnup}), we obtain:
\begin{eqnarray}\label{delG-1}
\delta (\Gamma^{-1}) ({\bf K},\Omega)=-\sum_{\bf k} \frac{\Sigma \us ({\bf k},\Omega+\mu ^0\ds-E({\bf K}-{\bf k}) )}
{(\Omega+\mu\us+\mu ^0\ds-\epsilon _k-E({\bf K}-{\bf k}))^2}-\frac{d}{dx}\sum_{\bf k}
\frac{ \Sigma \us ({\bf k},\epsilon_k-x )\theta(x-\epsilon_k)}{\Omega +\mu ^0\ds-\epsilon_k-E({\bf K}-{\bf k})+x} \;\Big|_{x=\mu \us}
\end{eqnarray}
One can check on this expression that $\delta (\Gamma^{-1}) ({\bf K},\Omega)$, and accordingly $\Gamma ({\bf K},\Omega)$,
is as expected analytical for ${\rm Re}\,\Omega <0$. Indeed the apparent (double) pole at $\Omega=\epsilon _k+E({\bf K}-{\bf k})
-\mu\us-\mu ^0\ds$ does not exist since one can check that the numerators combine exactly to leave an analytical function
at this point. This property allows to close again in Eq.(\ref{formud}) the contour around the negative frequency half-plane.
From Eq.(\ref{Gup}) this leaves us to evaluate the contribution coming from the poles of $G_{0\up}$ and $\delta G_{\up}$,
this last one being again handled as in Eq.(\ref{deltnup}).

In obtaining the first order variation $\delta \mu \ds$ of $\mu \ds$ from Eq.(\ref{formud}), one last point has to be taken into account.
Since there is a variation of $\mu \ds$ and that, in the zeroth order contribution Eq.(\ref{formud0}), $ \Gamma_0$ itself depends
on $\mu \ds$ as seen in Eq.(\ref{defG0}), we have to take into account this variation. In other words we have to keep in mind that
Eq.(\ref{formud}) is an implicit equation for $\mu \ds$. This  introduces again the coefficient $Z$ which has been introduced in Eq.(\ref{defZ0})
in the preceding subsection:
\begin{eqnarray}\label{eqZ}
Z=1-\sum_{\bf K} \theta(\mu \us-\epsilon_K) \frac{\partial}{\partial \Omega }\Gamma_0({\bf K},\epsilon_K-\mu \us)
=1+\sum_{\bf k K}\theta(\mu \us-\epsilon_K) \theta(\epsilon_k-\mu\us)
 \frac{\Gamma^2_0({\bf K},\epsilon_K-\mu \us)}{[E^{(1)}_{{\bf k}{\bf K}}]^2}
\end{eqnarray}
where we have used the shortened version $E^{(1)}_{{\bf k}{\bf K}} \equiv E^{(1)}_{{\bf 0}{\bf 0}{\bf k}{\bf K}}$ of our notation Eq.(\ref{defE1}). This quantity is actually just $1-\partial \Sigma\ds / \partial \omega $, that is the inverse of the residue of $G\ds$ at its pole for zero momentum and energy.

Making use of all these ingredients in Eq.(\ref{formud}), together with $\delta \Gamma = -\Gamma^2_0\,\delta (\Gamma^{-1})$, we obtain for the variation $\delta \mu \ds$ of $\mu \ds$ to the first order in $n\ds$:
\begin{eqnarray}\label{deltmudgdcan}
Z\frac{\delta \mu \ds}{n\ds}&=& \sum_{\bf k K}\frac{\theta(\mu \us-\epsilon_K)\Gamma^2_0({\bf K},\epsilon_K-\mu \us)
\Gamma_0({\bf k},\epsilon_K-E({\bf K}-{\bf k})-\mu \us+\mu ^0\ds)}{[E^{(1)}_{{\bf k}{\bf K}}]^2}
-\frac{d}{dx}\sum_{\bf K}\theta(x-\epsilon_K)\Gamma^2_0({\bf K},\epsilon_K-x)\;\Big|_{x=\mu \us} \\ \nonumber
&+&\sum_{\bf K}\theta(\mu \us-\epsilon_K)\Gamma^2_0({\bf K},\epsilon_K-\mu \us)\frac{d}{dx}\sum_{\bf k}
\frac{\theta(x-\epsilon_k)\Gamma_0({\bf k},\epsilon_k-x)}{\epsilon_K-\epsilon_k-E({\bf K}-{\bf k})
-\mu \us+\mu ^0\ds+x}\;\Big|_{x=\mu \us}
\end{eqnarray}

However this is not yet the result we are looking for. Indeed this variation is at fixed $\mu \us$ whereas, just
as in the preceding parts, we want to work in the canonical ensemble and find $\partial \mu \ds /\partial n\ds$
at fixed $n\us$, while Eq.(\ref{deltmudgdcan}) gives us $\partial \mu \ds/\partial n\ds$ at fixed $\mu \us$.
These two quantities are related by:
\begin{eqnarray}\label{gdcan}
\left(\frac{\partial \mu \ds}{\partial n\ds}\right)_{n \us}=\left(\frac{\partial \mu \ds}{\partial n\ds}\right)_{\mu  \us}
-\left(\frac{\partial \mu \ds}{\partial n\us}\right)_{n  \ds} \left(\frac{\partial n \us}{\partial n\ds}\right)_{\mu  \us}
\end{eqnarray}
From Eq.(\ref{self1}) and Eq.(\ref{deltnup1}) we have:
\begin{eqnarray}\label{deltnup2}
 \left(\frac{\partial n \us}{\partial n\ds}\right)_{\mu  \us}
= -\sum_{\bf k K}\theta(\mu \us-\epsilon_K) \theta(\epsilon_k-\mu\us)
 \frac{\Gamma^2_0({\bf K},\epsilon_K-\mu \us)}{[E^{(1)}_{{\bf k}{\bf K}}]^2}
-N_F \Gamma_0 (k_F,0) \\ \nonumber
\end{eqnarray}
while, since we are in the limit of a vanishing number of polarons $n\ds\rightarrow0$, 
$\partial \mu \ds/\partial n\us$ is merely obtained from the lowest order result Eq.(\ref{formud0}) for the polaron
chemical potential:
\begin{eqnarray}\label{mdmu}
ZN_F\left(\frac{\partial \mu \ds}{\partial n\us}\right)_{n  \ds=0}=Z\,\frac{d\mu ^0 \ds}{d\mu \us}=
N_F \Gamma_0(k_F,0)+\sum_{\bf k K}\frac{\theta(\mu \us-\epsilon_K)\delta(\epsilon _k-\mu \us)
\Gamma^2_0({\bf K},\epsilon_K-\mu \us)}{E^{(1)}_{{\bf k}{\bf K}}}
\end{eqnarray}
where $N_F= d\mu \us/dn\us= \sum_{k}\delta(\epsilon_k-\mu \us)$ is the density of states at the spin-up Fermi
level, and the factor $1/Z$ takes again into account that $\Gamma_0$ depends on $\mu \ds$. Our result is
obtained from Eq.(\ref{deltmudgdcan}),Eq.(\ref{gdcan}),Eq.(\ref{deltnup2}) and Eq.(\ref{mdmu}).

It is interesting to consider the weak coupling limit which provides a consistency check of our complicated
result. Indeed in this case we have $\Gamma_0=2\pi a/m_r$ and,
as we have seen in Eq.(\ref{selfwc}), $\Sigma\us({\bf k},\omega)$ is a constant 
which gives just a shift of the spin-up chemical potential. We have merely to replace $\mu \us$ by
$\mu \us-2\pi a n\us/m_r$. Otherwise the situation is not modified for the
spin-up particles which behave still as non-interacting particles. Hence the situation is unchanged with
respect to the calculation Eq.(\ref{formud0}) to lowest order in $n\ds$. However, in the canonical ensemble, 
we are interested in the dependence of the polaron chemical potential $\mu \ds$ on the spin-up particle 
number $n\us$. Accordingly the formula giving $\mu \ds$ in terms of $n\us$, i.e. in terms of $1/k_Fa$, is
unchanged in this weak coupling regime, and we should not find any correction. 

Indeed in this case we have \cite{crlc} $\mu \ds=2\pi n\us a/m_r$ and $\partial \mu \ds/\partial n\us=\Gamma_0$
in agreement with Eq.(\ref{mdmu}), as it should be, since from Eq.(\ref{eqZ}), we have $Z \simeq 1$,
the correction being of second order in $\Gamma_0$, i.e. in $a^2$. 
In Eq.(\ref{deltnup2}), the first term, which is of order $a^2$, is negligible compared to the second one
of order $a$, leading to $\partial n \us/\partial n\ds=-N_F \Gamma_0$.
Hence the second term in the right-hand side of Eq.(\ref{gdcan}) is $N_F \Gamma^2_0$. On the other hand
the first and third terms in the right-hand side of Eq.(\ref{deltmudgdcan}) are proportional to $\Gamma^3_0$,
i.e. to $a^3$. Hence in the weak coupling limit they are negligible compared to the second one which is
proportional to $\Gamma^2_0$. Again we have only to take in this term the derivative of the Heaviside
function and we find in this limit $\partial \mu  \ds/\partial n\ds
\Big|_{\mu \down}=-N_F \Gamma^2_0$. As a result, when we carry this value into Eq.(\ref{gdcan}), we
find that $\partial \mu  \ds/\partial n\ds\Big|_{n \up}$, which is formally of order $a^2$ in our calculation,
is actually exactly zero, as it should be.

Similar simplifications are also present in the general case, and we can write the final result explicitly as:
\begin{eqnarray}\label{eqZmn}
Z\,\left(\frac{\partial \mu \ds}{\partial n\ds}\right)_{n \us}&=& \sum_{\bf p q}\frac{\Gamma^2_0({\bf q},\omega _{\bf q})
\Gamma_0({\bf p},\omega _{\bf p q})}{[E^{(1)}_{{\bf p}{\bf q}}]^2}
-\sum_{\bf q q'}\frac{\Gamma^2_0({\bf q},\omega _{\bf q})
\Gamma_0({\bf q'},\omega _{\bf q'})}{[E^{(1)}_{{\bf q'}{\bf q}}]^2}
-2 \sum_{\bf k q}\frac{\Gamma^3_0({\bf q},\omega _{\bf q})}{[E^{(1)}_{{\bf k}{\bf q}}]^2} \\ \nonumber
&-&\sum_{\bf k q q'}
\frac{\Gamma^2_0({\bf q},\omega _{\bf q})\Gamma_0^2({\bf q'},\omega _{\bf q'})}{E^{(1)}_{{\bf q'}{\bf q}}[E^{(1)}_{{\bf k}{\bf q'}}]^2}
+\Gamma_0(k_F,0)\sum_{\bf k q}\frac{\Gamma^2_0({\bf q},\omega _{\bf q})}{[E^{(1)}_{{\bf k}{\bf q}}]^2} 
+\sum_{\bf k q q'}\frac{\Gamma^2_0({\bf q},\omega _{\bf q})\Gamma_0^2({\bf q'},\omega _{\bf q'})}{[E^{(1)}_{{\bf k}{\bf q}}]^2}\langle\frac{1}{E^{(1)}_{{\bf k_F}{\bf q'}}}\rangle
\end{eqnarray}
Here we have used our implicit notations $k>k_F$ and $q<k_F$, $q'<k_F$, while  ${\bf p}$ runs over all possible values of the wavevector. 
We have set $\omega _{\bf q}=\epsilon _{\bf q}-\mu \us$, and $\omega _{\bf p q}=\omega _{\bf q}-E({\bf p}-{\bf q})+\mu ^0\ds$.
Finally, in the last term, the brackets denote the angular average on the direction of ${\bf k}$ running on the Fermi surface $\epsilon_k=\mu \us$,
which is easily performed analytically. Note that the apparent singularities in the first two terms, which would occur for $E^{(1)}_{{\bf q'}{\bf q}}=0$,
are actually not present because the numerators of these first two terms cancel at the same place.

\begin{figure}
\includegraphics[width=\linewidth]{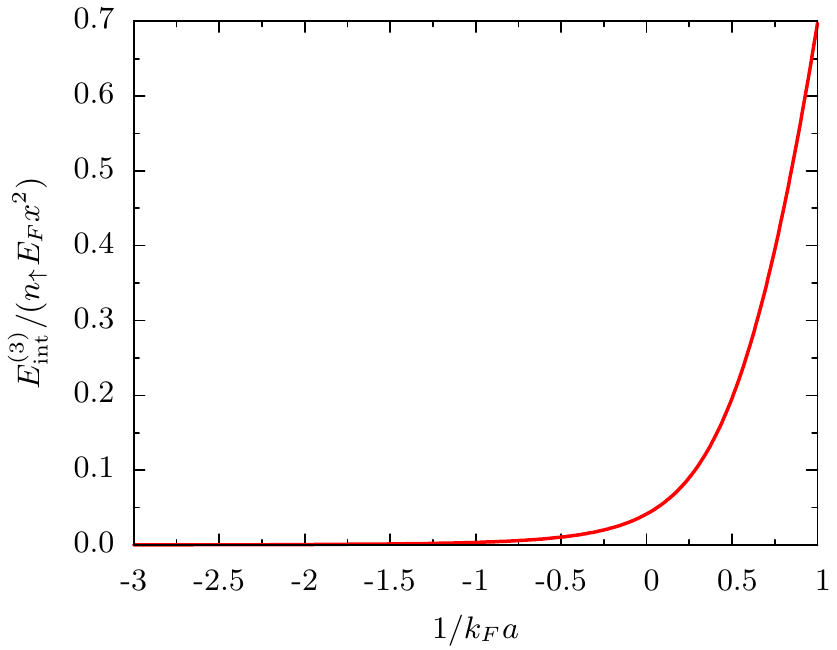}
\caption{\label{fig3} (Color online) $E_{\rm int}^{(3)}/(n\us E_F x^{2})$ as a function of
$1/k_F a$ in the case of equal masses $m\us=m\ds$.}
\end{figure}

Once the right-hand side of Eq.(\ref{eqZmn}) is evaluated, the corresponding contribution $E_{\rm int}^{(3)}$ to the interaction energy is given by:
\begin{eqnarray}\label{eint3}
\frac{E_{\rm int}^{(3)}}{n\us E_F}=\frac{1}{2}\frac{n\us}{E_F}\,\left(\frac{\partial \mu \ds}{\partial n\ds}\right)_{n \us}\,x^2
\end{eqnarray}
The coefficient in front of $x^{2}$ in Eq.(\ref{eint3}) is plotted in Fig.\ref{fig3}
for equal masses $m\us=m\ds$. It is worth noting that the result is positive, a result expected physically but by no means
obvious from the explicit expression Eq.(\ref{eqZmn}). It should be noted that the result is quantitatively fairly small. 
For example at unitarity the coefficient is 0.041.
This can be understood by noticing that the result goes very rapidly to zero in the weak coupling limit since we have seen that
it behaves as $a^3$ in this limit.

\section{Discussion and conclusion}

In the preceding sections we have calculated contributions to the expansion in powers of the density $x$ of the total system energy.
They correspond to the $x^2$ term and the $x^{7/3}$ term in this expansion. We note that this expansion is not restricted to integral powers
$x^n$ as one would expect classically, but rather to integral powers $x^{n/3}$ of $x^{1/3}$, that is we have actually an expansion in integral
powers of the polaron Fermi sea radius $p_F$. Naturally a few terms are missing at the beginning of this expansion and the first terms are
the $x$ term, coming from the polaron binding energy, and the $x^{5/3}$ term, coming from the polaron kinetic energy. We have found terms
proportional to $x^2$ and $x^{7/3}$, and it is quite likely that all the higher powers of $x^{1/3}$ have nonzero coefficients. For example
the polaron effective mass depends also on the polaron density, and we expect the first correction to the $x=0$ result to be proportional
to $x$ (presumably corresponding to a decrease of the effective mass). This will lead to a $x^{8/3}$ term (presumably with a positive coefficient). 

It should be noted that we have not proceeded to a systematic expansion in powers of $x$ within a single coherent framework. 
Proceeding in such a way would require a diagrammatic analysis. This is certainly a desirable goal, but unfortunately this looks fairly 
complicated to perform effectively. Our position has rather been to start from a physical point of view, and then to calculate each 
contribution in the more convenient way, either by the simple hamiltonian approach or diagrammatically. Nevertheless it
is clear that, in such a diagrammatic analysis, the term calculated in section \ref{polarFermup} will appear as a polaron-polaron interaction term
mediated by the spin-up Fermi sea (while the one calculated in section \ref{polarFerm} is just a direct manifestation of the polaron
Fermi sea). Our lack of systematic expansion leaves open in principle the possibility that, up to the order we have considered,
some terms are missing, although we believe that this is quite unlikely.

\begin{figure}
\includegraphics[width=\linewidth]{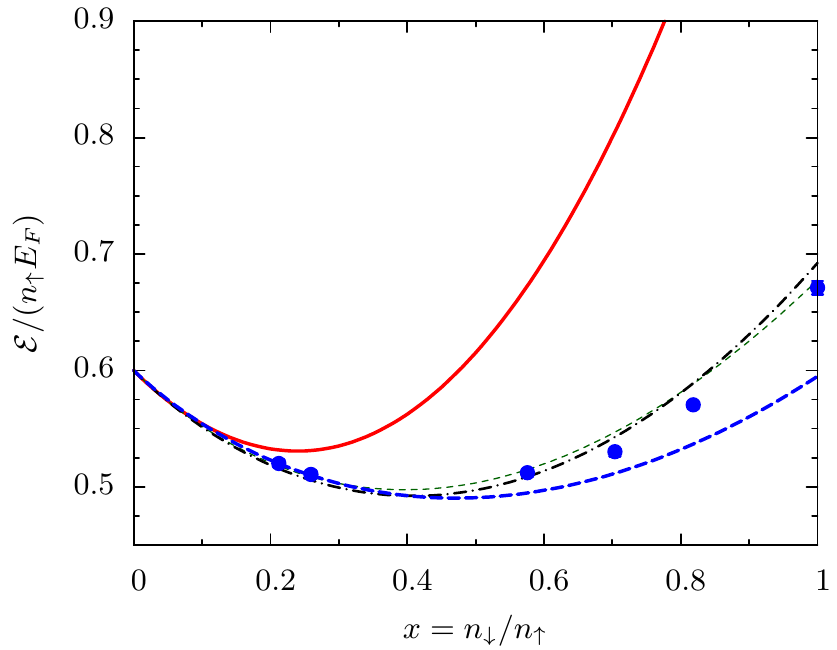}
\caption{\label{fig4} (Color online) Total reduced energy ${\mathcal E}/(n\us E_F)$ at unitarity as a function of the relative spin-down population
$x=n\ds / n\us$ (note that in Ref.\cite{lrgs} this is $5{\mathcal E}/(3n\us E_F)$ which is plotted). The (blue) thick dashed line is the expansion in powers 
of $x$ limited to the $x^{5/3}$ term with the coefficients from
Ref.\cite{cg}, i.e. the three first terms of Eq.(\ref{toten}). 
The (red) full line is this expansion going now up to the $x^{7/3}$ term, i.e. our full result Eq.(\ref{toten}).
The (black) dotted-dashed line is the same expansion, 
but omitting the dominant contribution coming from $E_{\rm int}^{(1)}$, and retaining only $E_{\rm int}^{(2)}$ and $E_{\rm int}^{(3)}$.
The (green) thin dashed line is the phenomenological formula Eq.(\ref{phen}) with $p=1.2$.
The blue dots are the Monte-Carlo results from Ref.\cite{lrgs}.}
\end{figure}

We can now gather our results to get the total energy ${\mathcal E}$ out of our results. We have to add to our interaction energy terms
the energy of the free spin-up Fermi sea, the contribution from the chemical potential of isolated polarons and the kinetic
energy of the non interacting polaron Fermi sea \cite{lrgs}. Focusing on the specific case of unitarity as in Ref.\cite{lrgs},
we have:
\begin{eqnarray}\label{toten}
\frac{\mathcal E}{n\us E_F}=\frac{3}{5}-0.6156\,x+\frac{3}{5}\frac{x^{5/3}}{1.20}+\frac{E_{\rm int}^{(1)}+E_{\rm int}^{(2)}+E_{\rm int}^{(3)}}{n\us E_F}
\end{eqnarray}
where the polaron binding energy and effective mass are taken from \cite{cg}. The result is displayed in Fig.~\ref{fig4} and it is in clear disagreement with
the Monte-Carlo results and the analytical curve of Ref.\cite{lrgs} as soon as $x \gtrsim 0.1$. Naturally in this range it is reasonable to believe the numbers given 
by Monte-Carlo calculations since they take full account of all the effects coming from the polaron finite density (although they work with a small
number of particles while in contrast our calculations are in the thermodynamical limit).

The most natural way to understand this discrepancy is to remark that there is no reason to believe that, in our powers expansion, terms of higher
order than the ones we have considered (i.e. going beyond $x^{7/3}$) do not contribute in a very important way. This is supported by the fact that
the contribution from $E_{\rm int}^{(1)}$ is quite important. This finding implies that the series is not rapidly converging and that accordingly terms
beyond this one give also an important contribution. Actually, at the start of the calculation, the only hope of agreement between our expansion 
and Monte-Carlo results was to find the opposite situation, that is all the terms we would find would be small. This would imply a rapidly converging 
series expansion which could be stopped at the order we have reached. Nevertheless this would have meant that polarons are weakly interacting
objects, which is hard to understand physically since obviously the polarization cloud is easily perturbed. Hence, although our result is disappointing
for quantitative agreement, it is quite reasonable physically. In this respect we note that, in Ref.\cite{lrgs}, the reasonable agreement between Monte-Carlo
results and the analytical result corresponding to the first three terms in Eq.(\ref{toten}) was to a large extent 
due to the use of the value $m^*/m\us \simeq 1$ (obtained
from Monte-Carlo calculations) for the polaron effective mass. Surprisingly the result is quite sensitive to the value of $m^*$, and making use
of $m^*/m\us= 1.20$ leads to an analytical result fairly different from the Monte-Carlo results. This sizeable difference can quite naturally be
interpreted as the important overall effect of the interaction between polarons. We note incidentally that, if we omit the dominant interaction
contribution $E_{\rm int}^{(1)}$, and retain only $E_{\rm int}^{(2)}$ and $E_{\rm int}^{(3)}$, we obtain a result in very good agreement with
Monte-Carlo calculations as it can be seen on Fig.\ref{fig4}.

There is naturally another possible source for the discrepancy we have found. Our calculation of the various coefficients is only approximate
since we have stayed within the single particle-hole approximation. For the calculation of the polaron binding energy and effective mass,
this approximation has proved to be remarkably accurate because taking more particle-hole excitations leads to a series which converges
extremely rapidly toward the exact result \cite{cg}. It is likely that a similar conclusion applies for the coefficients we have calculated. This
was our reason, in addition to simplicity, to stay at this level of approximation. However there is no good reason to believe that, if convergence
there is, it is as fast as for the binding energy. The effective mass displays already a somewhat slower convergence. Hence it is quite
possible that the convergence is not as fast for our coefficients. In particular we have noticed, when calculating $E_{\rm int}^{(1)}$, that
keeping only a single particle-hole excitation was leading to quite inconvenient situations. Even if we have found our way around these
problems, it seems likely they indicate that the single particle-hole approximation is not so accurate. Hence it is quite possible that the
exact results for the power expansion coefficients are somewhat away from our findings. Nevertheless we believe it is quite unlikely that
the corresponding changes reduce the interaction effects to such an extent that the polaron-polaron interaction could be considered as 
small enough to be neglected.

We consider now the work of Mora and Chevy \cite{chrisfred} in the light of our results. First of all they have neglected the direct interaction
term we have calculated in section \ref{polar} since it is a $x^{7/3}$ term, which is a coherent point of view. Nevertheless, as we have stressed,
this term becomes negligible only for very small $x$, in a range which is below the one involved in experiments or in Monte-Carlo calculations.
We are then left with the $x^2$ term. Here we identify the term they have found with the term, imposed by thermodynamics, to go from the grand 
canonical to the canonical ensemble, i.e. the second term in the right-hand side of Eq.(\ref{gdcan}). Indeed, just as in Eq.(\ref{eint3}), this term
gives to the total energy a contribution:
\begin{eqnarray}\label{compar}
{\mathcal E}_{\rm therm}=-\frac{1}{2}n\us^2\,\left(\frac{\partial \mu \ds}{\partial n\us}\right)_{n  \ds} \left(\frac{\partial n \us}{\partial n\ds}\right)_{\mu  \us}\,x^2
\end{eqnarray}
Since, as we have seen, thermodynamics implies $\partial n\us /\partial n\ds \big|_{\mu  \us}=- \partial \mu \ds /\partial \mu \us \big|_{n\ds}$,
and $\partial \mu \ds / \partial n\us\big|_{n  \ds}=(2\mu \us/3 n\us)\partial \mu \ds / \partial \mu \us\big|_{n  \ds}$, we obtain:
\begin{eqnarray}\label{compar}
{\mathcal E}_{\rm therm}=\left[\frac{3}{5}n\us \mu \us \right]\,\frac{5}{9}\,\left(\frac{\partial \mu \ds}{\partial \mu \us}\right)^2_{n  \ds}\,x^2
\end{eqnarray}
which is just their result for the interaction term. This is coherent with their assumption that in the grand canonical ensemble, one has a mixture
of two ideal Fermi gases of polarons and majority atoms, which means that there is no term describing their interaction.
Hence in the canonical ensemble only the term required by thermodynamics appears. By contrast we have found a contribution in the grand
canonical ensemble, namely the first term in Eq.(\ref{gdcan}), and in the weak coupling limit $a \to 0$ we find an overall result which behaves
like $a^3$ in contrast with their $a^2$ behaviour. Finally they ascribe their result to the effect of Pauli blocking by the polaron Fermi sea which
they find behaving as $x^2$. It is tempting to identify this effect with the one we have calculated in section \ref{polarFerm}, but we have found
a $x^{7/3}$ behaviour which is negligible in their framework. Finally we have mentioned that our $x^2$ term can be seen as an indirect
interaction between spin-down atoms mediated by the spin-up Fermi sea. This is in agreement with the interpretation proposed by
Yu, Z\"ollner and Pethick \cite{yzp}. However our microscopic result Eq.(\ref{eqZmn}) is much more complex than the one they propose
within Fermi liquid theory. This might be due to the fact that, in this way, they restrict themselves to indirect interaction through low
frequency perturbations, while we have not assumed such a restriction. It is also puzzling that they end up with the result obtained
by Mora and Chevy, while we have seen that it has a purely thermodynamical interpretation.

Finally it is interesting to present a simple phenomenological model which displays explicitly the troubles one meets when trying
to perform a power expansion. In this model we treat at first symmetrically the spin-up and spin-down atoms. First there is the standard
kinetic energy associated with the respective Fermi seas. Concentrating on the unitary case (and assuming also equal masses), 
we can in a simple phenomenology
disregard the difference between the polaron effective mass and the bare mass. Introducing the total density $n=n\us+n\ds$
and the associated Fermi energy $E_F^0=(6\pi ^2 n)^{2/3}/2m$, the sum of the kinetic energies of the two Fermi seas is
$nE_F^0\,(3/5)(u^{5/3}+v^{5/3})$ where $u=n\us/n$ and $v=n\ds/n$ (and $u+v=1$). With respect to the polaron binding energy
we know its value for $n\ds \to 0$ (or $n\us \to 0$). However the effective binding energy decreases when $n\ds$ increases
because the polarization of the spin-up Fermi sea has to be shared between all the spin-down atoms (this is another way to
see the polaron-polaron interaction). On the other hand this binding energy will clearly be zero when $n\ds =n$ since 
there is no spin-up Fermi sea anymore. Phenomenologically we may choose an interpolating function between these two
limits, and write for example the interaction energy as $-n\ds E_F\,\epsilon _b(v)=-n E_F^0\,u^{2/3}v\epsilon _b(v)$ with $\epsilon _b(v)=0.6\,(1-v)^p$, 
allowing for one fitting parameter $p$. Writing the corresponding expression for the spin-up polarons, we end up with
the following expression for the total energy ${\mathcal E}$
\begin{eqnarray}\label{}
\frac{{\mathcal E}}{nE_F^0}=\frac{3}{5}(u^{5/3}+v^{5/3})-u^{2/3}v\epsilon _b(v)-v^{2/3}u\epsilon _b(u)
\end{eqnarray}
The corresponding expression for ${\mathcal E}/n\us E_F$ in terms of $x=n\ds /n\us$ is obtained by dividing the above
expression by $u^{5/3}$, and using $u=1/(1+x)$ and $v=x/(1+x)$. This leads to:
\begin{eqnarray}\label{phen}
\frac{{\mathcal E}}{n\us E_F}=\frac{3}{5}(1+x^{5/3})-x\epsilon _b\left(\frac{x}{1+x}\right)-x^{2/3}\epsilon _b\left(\frac{1}{1+x}\right)
\end{eqnarray}
It turns out that $p=1.2$ gives a very good fit to the Monte-Carlo results as it can be seen from Fig.\ref{fig4}. On the other hand
since the radius of convergence for the series expansion of $1/(1+x)$ in powers of $x$ is 1, the expansion of ${\mathcal E}$
in powers of $x$ diverges for $x=1$. Hence there is no way with a power expansion to have an agreement with Monte-Carlo 
results for $x=1$. And accordingly this cast some doubts on the possibility to get good results for lower values of $x$
by making a series expansion and retaining a fairly large number of terms.

\section{acknowledgements}

We acknowledge stimulating discussions with F. Chevy, X. Leyronas and C. Mora. The ``Laboratoire de
Physique Statistique'' is ``Laboratoire associ\'e au Centre National de la Recherche
Scientifique et aux Universit\'es Paris 6 et Paris 7''. S.G. acknowledges the support of
the Alexander von Humboldt foundation for this work.

\end{document}